\pdfoutput=1

\documentclass[twocolumn,english]{revtex4}
\usepackage{times,amssymb,amsmath,graphicx}
\usepackage[bookmarks=false,pdffitwindow=false,pdfstartview={FitH}]{hyperref}
\usepackage[T1]{fontenc}
\usepackage{babel}
\usepackage{color}

\begin{document}

%%  ======================================================================  %%
%%  TO JEST  >> W A Z <<  O DLUGOSCI 70+8 ZNAKOW I MA SIE MIESCIC W LINII   %%
%%  ======================================================================  %%

\title{
  Theory of the 
  sub-Sharvin charge transport in graphene disks
}

\author{Adam Rycerz}
\affiliation{Institute for Theoretical Physics,
  Jagiellonian University, \L{}ojasiewicza 11, PL--30348 Krak\'{o}w, Poland}

\author{Piotr Witkowski}
\affiliation{Institute for Theoretical Physics,
  Jagiellonian University, \L{}ojasiewicza 11, PL--30348 Krak\'{o}w, Poland}

\date{September 28, 2022}

\begin{abstract}
  Ballistic graphene samples in a~multimode regime show the sub-Sharvin
  charge transport, characterized by the conductance reduced by a~factor of
  $\pi/4$ comparing to standard Sharvin contacts in two-dimensional electron
  gas, and the shot-noise power enhanced up to $F\approx{}1/8$ (with $F$
  the Fano factor) [Phys.\ Rev.\ B {\bf 104}, 165413 (2021)].
  Here we consider the disk-shaped (Corbino) setup in graphene, with inner
  radius $r_1$ and outer radius $r_2$, finding that the multimode conductance
  is slightly enhanced for any $r_1<r_2$, reaching
  $(4\!-\!\pi)\approx{}0.8684$ of the Sharvin value for
  $r_1\ll{}r_2$. At the same limit, the Fano factor is reduced,
  approaching $(9\pi-28)/(12-3\pi)\approx{}0.1065<1/8$.
  Closed-form approximating expressions for any $r_1/r_2$ ratio are derived
  supposing
  incoherent scattering of Dirac fermions on asymmetric double barrier and
  compared with exact numerical results following from the mode-matching
  method. Sub-Sharvin values are restored in the narrow-disk limit
  $r_1/r_2\rightarrow{}1$. For experimentally-accessible radii ratios
  $0.5\leqslant{}r_1/r_2\leqslant{}0.8$ both the conductance and the Fano
  factor are noticeably closer to the values predicted for the
  $r_1\ll{}r_2$ limit, yet still differ from standard Sharvin
  transport characteristics.
  The system behavior upon tuning the electrostatic potential barrier from
  a~rectangular to parabolic shape is studied numerically, and the
  crossover from the sub-Sharvin to standard Sharvin transport regime is
  demonstrated.
  Implications for a~finite section of the disk are also discussed. 
\end{abstract}

\maketitle

\section{Introduction}
Several unique properties of graphene can be attributed to the fact that
material characteristics of this form of carbon are determined
by unusual properties of massless Dirac fermions in two dimensions
\cite{Sem84,Nov05,Zha05,Pet14,Zen19,Kuz08,Nai08,Kat06,Two06,Son08,Dan08,Lai16,
Bal08,Yos15,Cro16,Pal12,Ryc13,Hua18}.
In particular, chiral nature
of effective quasiparticles and conical dispersion \cite{Sem84,Nov05} lead
to half-integer sequence of quantum-Hall states \cite{Zha05,Pet14,Zen19}
and quantized light absorption \cite{Kuz08,Nai08}.
The phenomenon of Klein tunneling \cite{Kat06} and transport via evanescent
waves result in universal {\em dc} conductivity ($\sigma_0=4e^2/\pi{}h$, with
the electron charge $-e$ and the Planck constant $h$) and 
{\em pseudodiffusive} shot noise power (quantified by the Fano factor $F=1/3$)
\cite{Two06,Son08,Dan08,Lai16} in samples close to the charge-neutrality
point. Even though thermal conductivity of graphene is dominated by phonons
\cite{Bal08}, and therefore only partly-related to the properties of Dirac
electrons, some excess thermal conductance from these electrons \cite{Yos15},
violating the Wiedemann-Franz law obeyed by the Schr\"{o}dinger electrons,
was detected \cite{Cro16}. Also, the presence of the valley degree of freedom
affects several hallmarks of mesoscopic physics, including the conductance
and spectral fluctuations \cite{Pal12,Ryc13,Hua18}. 

Recently, it was shown using slightly different theoretical approaches
\cite{Par21,Ryc21a,Ryc21b} that the electrical conductance of a~rectangular
graphene sample away from the charge-neutrality point is reduced, while
the Fano factor in amplified comparing to standard ballistic systems
\cite{Naz09}, namely
\begin{equation}
  \label{subshar}
  G= \frac{\pi}{4}\,G_{\rm Sharvin}, \ \ \ \ \ \ F=\frac{1}{8},
\end{equation}
with the Sharvin conductance $G_{\rm Sharvin}=g_0k_FW/\pi$ being the upper bound
for the conductance of ballistic nanostructures \cite{Sha65,Bee91}.
The conductance quantum is $g_0=4e^2/h$ due to the spin and valley
degeneracies; the Fermi momentum $k_F$ is tuned using the gate electrode such
that $k_F^{-1}\ll{}\text{min}\left(W,L\right)$, with $W$ the sample width and
$L$ the sample length.
Although the sample conductance is rather difficult to determine
experimentally due to the resistances of contacts, existing experiments
report the Fano factor approaching $F\approx{}0.10\,\div\,{}0.15$
\cite{Dan08,Lai16} away from the charge-neutrality point, being significantly
greater than $F\approx{}0$ expected for ballistic systems. 

It is further found in Ref.\ \cite{Ryc21b} that the ballistic values of
$G\approx{}G_{\rm Sharvin}$ and $F\approx{}0$ are gradually restored when the
longitudinal potential barrier evolves from a~rectangular towards
a~parabolic shape. 

In the so-called {\em sub-Sharvin} transport regime, when $k_F$ is
approximately constant in the whole sample area (in other words, each
potential jumps in a~contact region occur on a~lengthscale
$\Delta{}x\ll{}\lambda_F/2$, with $\lambda_F$ being a~typical Fermi
wavelength in the sample area), transmission probability can be approximated
by \cite{Ryc21b}
\begin{equation}
  \label{ttrecappr}
  T\approx{}\cos\theta = \sqrt{1-\left({k_y}/{k_F}\right)^2}, 
\end{equation}
where $\theta$ denotes the direction of propagation in the central area
(with respect to the longitudinal axis) and $k_y$ is the transverse momentum
component. Summing over possible values of $|k_y|\leqslant{}k_F$ in accordance
with the Landauer-B\"{u}ttiker formula \cite{Lan57,But85}, one immediately
obtains the values given in Eq.\ (\ref{subshar}) provided that $k_FW\gg{}1$
and $k_FL\gg{}1$ \cite{kflenfoo}. 

\begin{figure*}[!t]
  \includegraphics[width=0.9\linewidth]{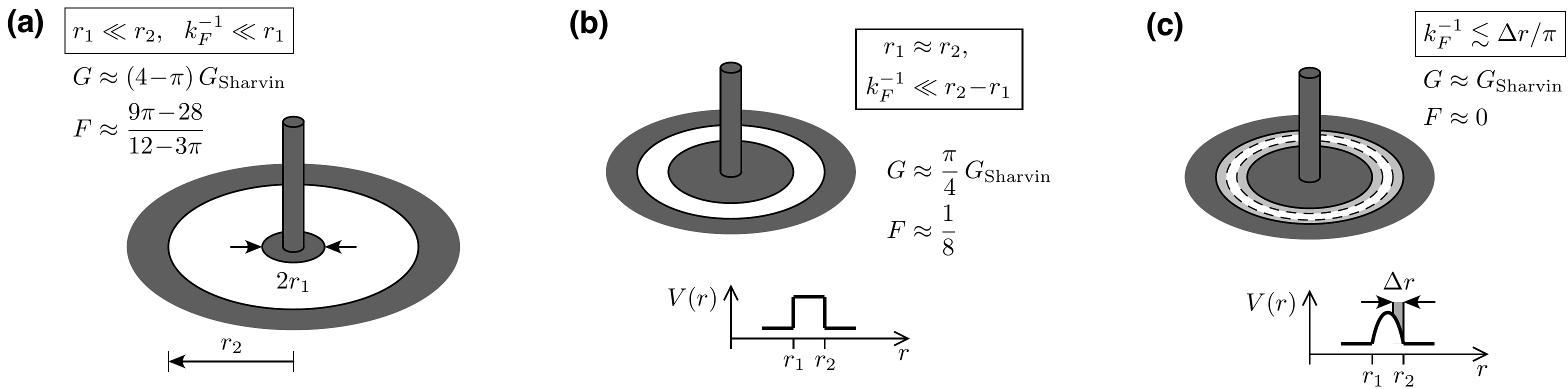}
  \caption{ \label{outline:fig}
    Outline of the results presented in the paper.
    For the Corbino disk, i.e., weakly-doped graphene annulus
    (white) attached to circular heavily-doped leads (dark), with
    the inner radius $r_1$ and the outer radius $r_2$, the conductance
    ($G$) and Fano factor ($F$) approaches, in the multimode regime
    (with the Fermi wavenumber satisfying $k_Fr_1\gg{}1$), the limiting
    values derived for different physical situations: 
    (a) For a~wide disk, $r_2\gg{}r_1$, the values for incoherent scattering
    of Dirac fermions on a~double barrier with perfect transmission on
    the outer potential step at the distance $r=r_2$ from the disk center,
    apply. 
    (b) For a~thin disk, $r_1\approx{}r_2$, the values for incoherent
    scattering on a~symmetric double barrier are restored. 
    (c) If the rectangular potential barrier is replaced with a~smooth one,
    with effective depth of a~potential step $\Delta{}r$,
    the standard Sharvin transport characteristics show up for
    $k_F\Delta{}r\gtrsim{}\pi$. 
  }
\end{figure*}

Remarkably, the above-mentioned result is insensitive to the system aspect
ratio ($W/L$) suggesting it may also be independent of the sample geometry.
However, the idealized boundary conditions used in theoretical considerations
usually lead to results which become comparable with the experiment starting
from $W/L\gtrsim{}20$, see Refs.\ \cite{Two06,Dan08}.
(For $W<L$, the conductance is suppressed, mainly due to the presence of edge
disorder \cite{Lin08,Tom11,Ihn12,Lib12}.) 
For these reasons, we consider here the edge-free Corbino geometry, for
which the existing experimental \cite{Pet14,Zen19,Kam21} and theoretical
\cite{Che06,Ryc09,Ryc10,Sus20} background allows to revisit the sub-Sharvin
charge-transfer characteristics in search for the geometry- (in particular, the
radii-ratio-) related effects which may be confirmed using existing devices. 
Main findings of the present work are summarized in Fig.\ \ref{outline:fig}.

The remaining parts of the paper are organized as follows.
In Sec.\ \ref{appcofan}, we derive an approximation for
the transmission through a~doped Corbino disk in graphene and subsequent
formulas for charge-transfer characteristics: the conductance and the Fano
factor. 
Comparison with the exact numerical results from the mode-matching analysis
is given in Sec.\ \ref{exadisk}.
Next, in Sec.\ \ref{smoopots}, we discuss the effects of tuning the potential
barrier from a~rectangular to a~parabolic shape.
The role of sample edges, modeled via the infinite-mass boundary
conditions, is studied in Sec.\ \ref{asecdisk}. 
The conclusions are given in Sec.\ \ref{conclu}.

\section{Approximate conductance and Fano factor for
  graphene disk}
\label{appcofan}

\begin{figure}[!b]
  \includegraphics[width=0.9\linewidth]{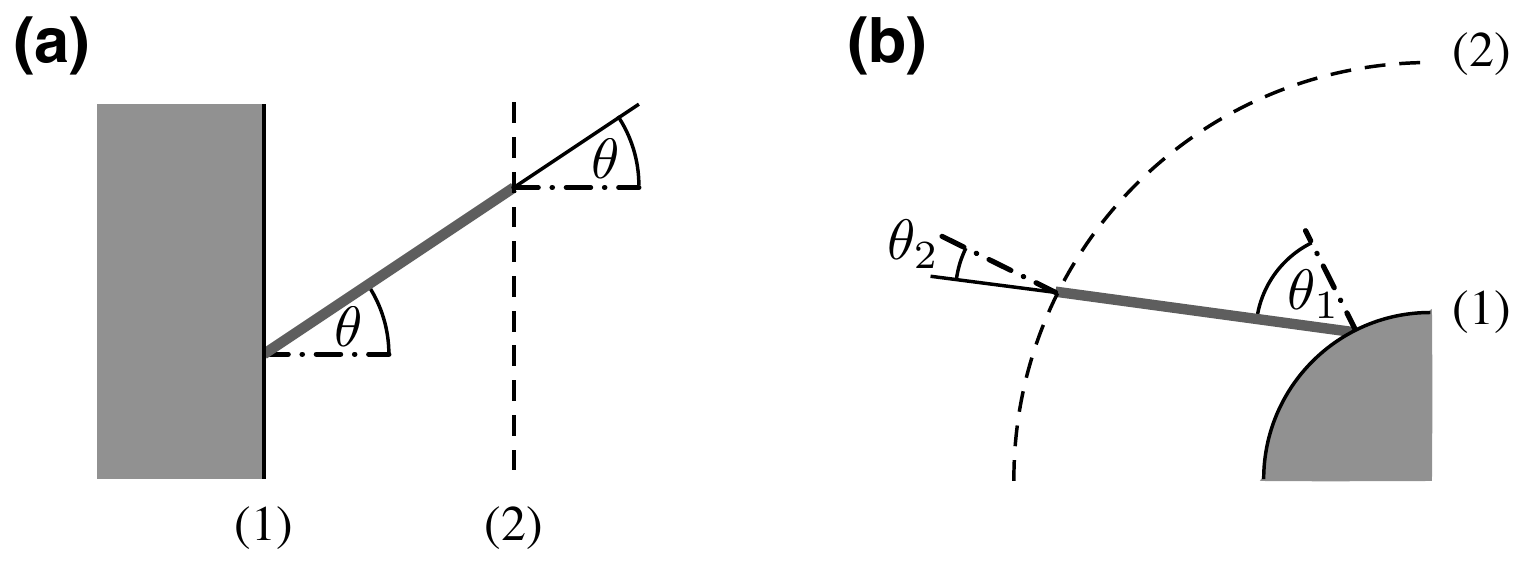}
  \caption{ \label{fig:scattering}
    Propagation between scattering on interfaces $(1)$ and $(2)$ separating
    weakly doped [white area] and heavily doped [shadow area; omitted for
    interface (2) for clarity] regions in graphene (schematic).
    (a) The rectangular geometry. (b) The Corbino disk. Angular momentum
    conservation implies that incident angles for a~particle
    satisfy $\theta_2<\theta_1$ (if $\theta_1>0$) for the disk case.
    (Notice that thin solid lines behind the interface (2), visually
    elongating the trajectory mark with thick lines, are guide for the eye
    only; same applies to dash-dotted lines marking the normal to an
    interface.) 
  }
\end{figure}

\subsection{Scattering on straight interfaces}
The reflection and transmission probabilities for the straight interface
separating weakly- and heavily-doped regions in graphene [see Fig.\
\ref{fig:scattering}(a)] can be written,
in the limit of an infinite doping on one side, as functions of the incident
angle $\theta$ on the other (i.e., weakly-doped) side \cite{Son08,Ryc09},
namely
\begin{equation}
  \label{ref1tra1}
  R_1 = \frac{1-\cos\theta}{1+\cos\theta}, \ \ \ \ \ \
  T_1 = \frac{2\cos\theta}{1+\cos\theta},  
\end{equation}
where $\cos\theta$ is related to the momentum component parallel to the
interface ($k_y$) via the second equality in Eq.\ (\ref{ttrecappr}).
In particular,
\begin{equation}
  \label{tra1ky}
  T_1 = \frac{2\,\sqrt{1-(k_y/k_F)^2}}{1+\sqrt{1-(k_y/k_F)^2}}. 
\end{equation}

For two interfaces in series, one can employ the double-contact formula
\cite{Dat97} for the transmission
\begin{equation}
  T = \frac{T_1T_2}{1+R_1{}R_2-2\sqrt{R_1{}R_2}\cos{\phi}}, 
\end{equation}
where $T_2$ and $R_2=1-T_2$ are transmission and reflection probabilities
for the second interface, and $\phi$ denotes the phase shift acquired during
a~single round-trip between the scatterers. 
Substituting $T_2=T_1$ as given by Eq.\ (\ref{tra1ky}) and
$\phi=2k_xL=2\sqrt{k_F^2-k_y^2}\,L$ for interfaces at a~distance $L$,
we obtain
\begin{equation}
  \label{ttkxky}
  T=\frac{1}{1+\left(k_y/k_x\right)^2\sin^2(k_xL)}. 
\end{equation}
The above holds true for propagating modes ($k_y\leqslant{}k_F$).
For evanescent modes ($k_y>k_F$) one can obtain the analytic continuation
by setting $k_x=i\sqrt{k_y^2-k_F^2}$. 

\subsection{Landauer-B\"{u}ttiker formalism}
For a~confined geometry, the quantization of $k_y$ appears. For instance,
if the infinite-mass confinement is assumed \cite{Ber87},
we have $k_y=k_y^{(n)}\equiv{}\pi(n+\frac{1}{2})/W$ with $n=0,1,2,\dots$
\cite{bcfoo}.
Substituting $k_y^{(n)}$ to Eq.\ (\ref{ttkxky}) we obtain the transmission
probability for $n$-th normal mode $T_n$. Next, the conductance and the
Fano factor follow by summing over the modes,
\begin{equation}
  \label{gfland}
  G = g_0\sum_{n=0}^{N-1}T_n, \ \ \ \ \ \ 
  F = \frac{\sum_{n=0}^{N-1}T_n(1-T_n)}{\sum_{n=0}^{N-1}T_n}, 
\end{equation}
where $N=\lfloor{}WK/\pi\rfloor$ is the number of propagating modes in the
leads for a~finite doping ($K$ denotes the Fermi momentum in the leads)
considered further in this paper. The limit of infinite doping corresponds
to $N\rightarrow\infty$ in Eq.\ (\ref{gfland}). 

Earlier in Ref.\ \cite{Ryc21b}, we have argued that for high doping
in the sample area, $k_F\gg{}W^{-1}$ and $k_F\gg{}L^{-1}$, the argument of sine
in Eq.\ (\ref{ttkxky}), i.e., $k_xL=\phi/2$, can be regarded as a~random phase
when summing contributions for consecutive $n$-s in Eq.\ (\ref{gfland}).
Also, the role of evanescent modes is negligible in such a~range.
In turn, $T_n$ can be approximated by substituting $k_y=k_y^{(n)}$ to the
rightmost formula in Eq.\ (\ref{ttrecappr}) for $k_y^{(n)}\leqslant{}k_F$,
or by $0$ for $k_y^{(n)}>k_F$. The corresponding formula for $(T_n)^2$ can be
derived by averaging a~square of Eq.\ (\ref{ttkxky}) over $\phi$, and reads
\begin{equation}
  \label{tt2recappr}
  (T_n)^2 \approx{} \sqrt{1-\left(\frac{k_y^{(n)}}{k_F}\right)^2}
  \left[1-\frac{1}{2}\left(\frac{k_y^{(n)}}{k_F}\right)^2\right].
\end{equation}
Additionally, since $k_F\gg{}\Delta{}k_y=\pi/W$ being the transverse
momentum quantum, the sums appearing in Eq.\ (\ref{gfland})
can be replaced by integrals over $0\leqslant{}k_y<k_F$,
leading to the expressions for $G$ and $F$ given in Eq.\ (\ref{subshar}). 

Here, we point out that the above-mentioned results can also be obtained
using the double-contact formula for incoherent transmission
\cite{Dat97,intfoo}, namely
\begin{align}
  \left\{T\right\}_{\rm incoh} &= \frac{1}{2\pi}\int_{-\pi}^{\pi}d\phi\, 
  \frac{T_1T_2}{1+R_1{}R_2-2\sqrt{R_1{}R_2}\cos{\phi}} \nonumber \\ 
  &= \frac{T_1T_2}{1-R_1R_2} =
  \frac{T_1T_2}{T_1+T_2-T_1T_2}.
  \label{tticoh}
\end{align}
Substituting $T_2=T_1$ given by Eq.\ (\ref{tra1ky}), we immediately obtain
relevant formulas in Eqs.\ (\ref{subshar}) and (\ref{ttrecappr}).
Similarly, calculating
\begin{align}
  \left\{T^2\right\}_{\rm incoh} &= \frac{1}{2\pi}\int_{-\pi}^{\pi}d\phi\,
  \left(\frac{T_1T_2}{1+R_1{}R_2-2\sqrt{R_1{}R_2}\cos{\phi}}\right)^2
  \nonumber \\
  &= \frac{(T_1T_2)^2(1+R_1R_2)}{(1-R_1R_2)^3} \label{tt2icoh}
\end{align}
brought us [for $T_2=T_1$ given by Eq.\ (\ref{tra1ky}), $R_{1(2)}=1-T_{1(2)}$,
again] to Eq.\ (\ref{tt2recappr}) and the value of $F$ given in
Eq.\ (\ref{subshar}). (Notice that evaluating incoherent square of
the transmission probability in Eq.\ (\ref{tt2icoh}), later used to
determine the shot-noise power, one need to calculate
squared coherent probability first, and then average the result over
a~random phase.)

\subsection{Implications for the Corbino disk}
Although the double-contact formula cannot be directly applied to the
Corbino disk [see Fig.\ \ref{fig:scattering}(b)] Eqs.\ (\ref{tticoh}) and
(\ref{tt2icoh}) give us a~useful tool to generate approximate formulas
for charge-transfer characteristics also in this case. 
Now, the angular momentum ($\hbar{}j$) is a~conserved quantity
for a~particle traveling through the disk area.
In turn, $k_y$ in Eq.\ (\ref{tra1ky}) needs to be replaced by $j/r_1$,
while an analogous expression for $T_2$ can be generated by substituting
$j/r_2$ instead of $k_y$ in Eq.\ (\ref{tra1ky}). Using the last formula
in Eq.\ (\ref{tticoh}) we obtain
\begin{equation}
  \label{ttdiskicoh}
  \left\{T\right\}_{\rm incoh} = \frac{2c_1c_2}{c_1+c_2},\ \ \text{with }\
  c_{1(2)}=\sqrt{1-\left(\frac{j}{r_{1(2)}k_F}\right)^2},  
\end{equation}
being related to incident angles in Fig.\ \ref{fig:scattering}(b) via
$c_{1(2)}=\cos\theta_{1(2)}$. Similarly, from Eq.\ (\ref{tt2icoh}) we get
\begin{equation}
  \label{tt2diskicoh}
  \left\{T^2\right\}_{\rm incoh} = 4\frac{c_1^2c_2^2(1+c_1c_2)}{(c_1+c_2)^3}. 
\end{equation}

In order to calculate measurable quantities we employ Eq.\ (\ref{gfland})
approximating (for $k_F\gg{}1/r_1$) sums by integrals over a~dimensionless
$u=j/(r_1k_F)$, in the interval $-1\leqslant{}u\leqslant{}1$.
For the conductance, we get 
\begin{align}
  G =& \,2g_0r_1k_F\int_0^1 du{}\,\left\{T\right\}_{\rm incoh} =
  G_{\rm Sharvin} \times \nonumber \\
  & 
  \frac{(2a+\frac{1}{a})\arcsin{}a+3\sqrt{1-a^2}-\frac{\pi}{2}(a^2+2)}{1-a^2},
  \label{ggdiskicoh}
\end{align}
where we have used parity of Eq.\ (\ref{ttdiskicoh}) upon
$j\leftrightarrow{}-j$ to shrink the integration range to
$0\leqslant{}u\leqslant{}1$, introduced $G_{\rm Sharvin}=2g_0r_1k_F$ being
the Sharvin conductance for a~disk, and defined the inverse radii ratio
$a=r_1/r_2<1$. The Fano factor now reads
\begin{widetext}
\begin{align}
  F &= 1 - \left.{\int_0^1 du{}\,\left\{T^2\right\}_{\rm incoh}}\right/
  \bigg({\int_0^1 du{}\,\left\{T\right\}_{\rm incoh}}\bigg)
  \nonumber \\
  &= \frac{2a\sqrt{1-a^2}(53+279a^2+88a^4)-3\pi{}a(12+82a^2+45a^4+a^6)+6(1+45a^2+82a^4+12a^6)\arcsin{}a}{6(1-a^2)^2\left[\pi{}a(a^2+2)-6a\sqrt{1-a^2}-2(2a^2+1)
  \arcsin{}a\right]}.
  \label{ffdiskicoh}
\end{align}
\end{widetext}

Asymptotic forms of Eqs.\ (\ref{ggdiskicoh}) and (\ref{ffdiskicoh}) are
the following
\begin{equation}
  \label{gfasym1}
  G\simeq{}\frac{\pi}{4}G_{\rm Sharvin}, \ \ \ \ F\simeq{}\frac{1}{8},
  \ \ \ \ \text{for }\ \ a\rightarrow{}1,
\end{equation}
and
\begin{equation}
\label{gfasym0} 
  G\simeq{}(4\!-\!\pi)\,G_{\rm Sharvin}, \ \ \ \
  F\simeq{}\frac{9\pi\!-\!28}{12\!-\!3\pi},
  \ \ \ \ \text{for }\ \ a\rightarrow{}0. 
\end{equation}
Formulas in Eq.\ (\ref{gfasym1}) refer to the thin-disk limit
($r_1\approx{}r_2$) and can be easily obtained by setting $c_2\approx{}c_1$ in
Eqs.\ (\ref{ttdiskicoh}) and (\ref{tt2diskicoh}) and repeating the subsequent
steps. Similarly, Eq.\ (\ref{gfasym0}) describes the wide-disk limit
($r_2\gg{}r_1$), and can be obtained after setting $c_2\approx{}1$ (being
equivalent to $T_2\approx{}1$ describing a~perfect transmission for normal
incidence) in Eqs.\ (\ref{ttdiskicoh}) and (\ref{tt2diskicoh}).
Therefore, in the $r_2\gg{}r_2$ limit the role of the outer interface
(at $r=r_2$) is suppressed, and transport properties of the system are
governed by scattering of Dirac fermions on a~single potential step at
the inner disk edge, at $r=r_1$.

\section{Exact solution for the disk}
\label{exadisk}

\subsection{Mode-matching for the Dirac equation}
The analysis starts from the Dirac equation for a~single valley ($K$),
which can be written as
\begin{equation}
  \label{direqvr}
  \left[
    v_F\,\mbox{\boldmath$p$}\cdot\mbox{\boldmath$\sigma$} + V(r)
  \right]
  \Psi=E\Psi, 
\end{equation}
where $v_F=\sqrt{3}\,t_0a/(2\hbar)\approx{}10^6\,$m$/$s is the
energy-independent Fermi velocity in graphene (with $t_0=2.7\,$eV
the nearest-neighbor hopping integral and $a=0.246$ the lattice parameter),
$\mbox{\boldmath$p$}=(p_x,p_y)$ is the in-plane momentum operator with
$p_j=-i\hbar{}\partial_j$, and $\mbox{\boldmath$\sigma$}=(\sigma_x,\sigma_y)$
with $\sigma_j$ being the Pauli matrices. 
Taking the wavefunction in polar coordinates in a~form
$\Psi_j(r,\varphi) = e^{i(j-1/2)\varphi}\left(\chi_a,\chi_be^{i\varphi}\right)^T$,
with $j=\pm{1/2},\pm{3/2},\dots$ the total angular-momentum
quantum number, brought us to the system of ordinary differential equations
for the spinor components
\begin{align}
  \chi_a' &= \frac{j-1/2}{r}\chi_a+i\frac{E-V(r)}{\hbar{}v_F}\chi_b,
  \label{phapri} \\
  \chi_b' &= i\frac{E-V(r)}{\hbar{}v_F}\chi_a-\frac{j+1/2}{r}\chi_b.
  \label{phbpri} 
\end{align}

In this section, our discussion is limited to a~piecewise-constant potential
energy $V(r)$, earlier considered in Ref.\ \cite{Ryc09}. 
For the electron-doping case, $E>V(r)$, solutions for
the incoming (i.e., propagating from $r=0$) and outgoing (propagating from
$r=\infty$) waves are given, up to the normalization, by
%\begin{align}
\begin{equation}
  \label{chijleads}
  \chi_j^{\rm in} \!= \left(
  \begin{array}{c}
    H_{j-1/2}^{(2)}(kr) \\
    iH_{j+1/2}^{(2)}(kr) \\
  \end{array}
  \right),
  % \\
  \ 
  \chi_j^{\rm out} \!= \left(
  \begin{array}{c}
    H_{j-1/2}^{(1)}(kr) \\
    iH_{j+1/2}^{(1)}(kr) \\
  \end{array}
  \right),
  \end{equation}
%\end{align}
where $H_\nu^{(1,2)}(\rho)$ is the Hankel
function of the (first, second) kind, and $k=|E-V(r)|/(\hbar{}v_F)$.
For the disk area, we have $V(r)=0$, and the solution can be represented as
\begin{equation}
  \label{chijdisk}
  \chi_j^{(d)}=A_j\chi_j^{\rm in}(k_Fr)+B_j\chi_j^{\rm out}(k_Fr), \ \ \ \ 
  r_1\!<\!r\!<\!r_2,
\end{equation}
where $A_j$ and $B_j$ are arbitrary constants, and the Fermi wavenumber 
$k_F=|E|/(\hbar{}v_F)$. 
For the hole doping case, $E<U(r)$, the wavefunctions are replaced by
$\tilde{\chi}_j^{\rm in(out)}=\left[\chi_j^{\rm in(out)}\right]^\star$, 
using the relation $H_\nu^{(2)}=\big[H_\nu^{(1)}\big]^\star$.

Heavily-doped graphene leads are usually modeled by taking the limit of
$V(r)=V_0\rightarrow{}\pm\infty$ for $r<r_1$ or $r>r_2$.
The corresponding wavefunctions simplify to
\begin{align}
  \chi_j^{(1)} &=
    \frac{e^{\pm{}iKr}}{\sqrt{r}}
    \left(\begin{array}{c} 1 \\ 1 \\ \end{array}\right)
    + r_j
    \frac{e^{\mp{}iKr}}{\sqrt{r}}
    \left(\begin{array}{c} 1 \\ -1 \\ \end{array}\right),
    \  &r<r_1,
  \label{chij1} \\
  \chi_j^{(2)} &= t_j
    \frac{e^{\pm{}iKr}}{\sqrt{r}}
    \left(\begin{array}{c} 1 \\ 1 \\ \end{array}\right),
    \  &r>r_2, \label{chij2} 
\end{align}
with the reflection (and transmission) amplitudes $r_j$
(and $t_j$) and $K=|E-V_0|/(\hbar{}v_F)\rightarrow{}\infty$.

Solving the mode-matching conditions, $\chi_j^{(1)}(r_1)=\chi_j^{(d)}(r_1)$
and $\chi_j^{(d)}(r_2)=\chi_j^{(2)}(r_2)$, 
we find the transmission probability for $j$-th mode
\begin{equation}
\label{tjphi}
  T_{j} = |t_j|^2 = \frac{16}{\pi^2{}k^2{}r_1{}r_2}\,
  \frac{1}{\left[\mathfrak{D}_{j}^{(+)}\right]^2
    + \left[\mathfrak{D}_{j}^{(-)}\right]^2},
\end{equation}
with
\begin{align}
\label{ddnupm}
  \mathfrak{D}_{j}^{(\pm)} = &\mbox{Im}\left[
    H_{j-1/2}^{(1)}(kr_1)H_{j\mp{}1/2}^{(2)}(kr_2)\right. \nonumber \\
    &\pm \left.H_{j+1/2}^{(1)}(kr_1)H_{j\pm{}1/2}^{(2)}(kr_2)
    \right]. 
\end{align}
The result given by Eqs.\ (\ref{tjphi}) and (\ref{ddnupm})
corresponds to rectangular shape and infinite height of the potential
barrier $V(r)$. Other potential barriers are considered
in Sec.\ \ref{smoopots}.

\begin{figure}[!t]
  \includegraphics[width=1.0\linewidth]{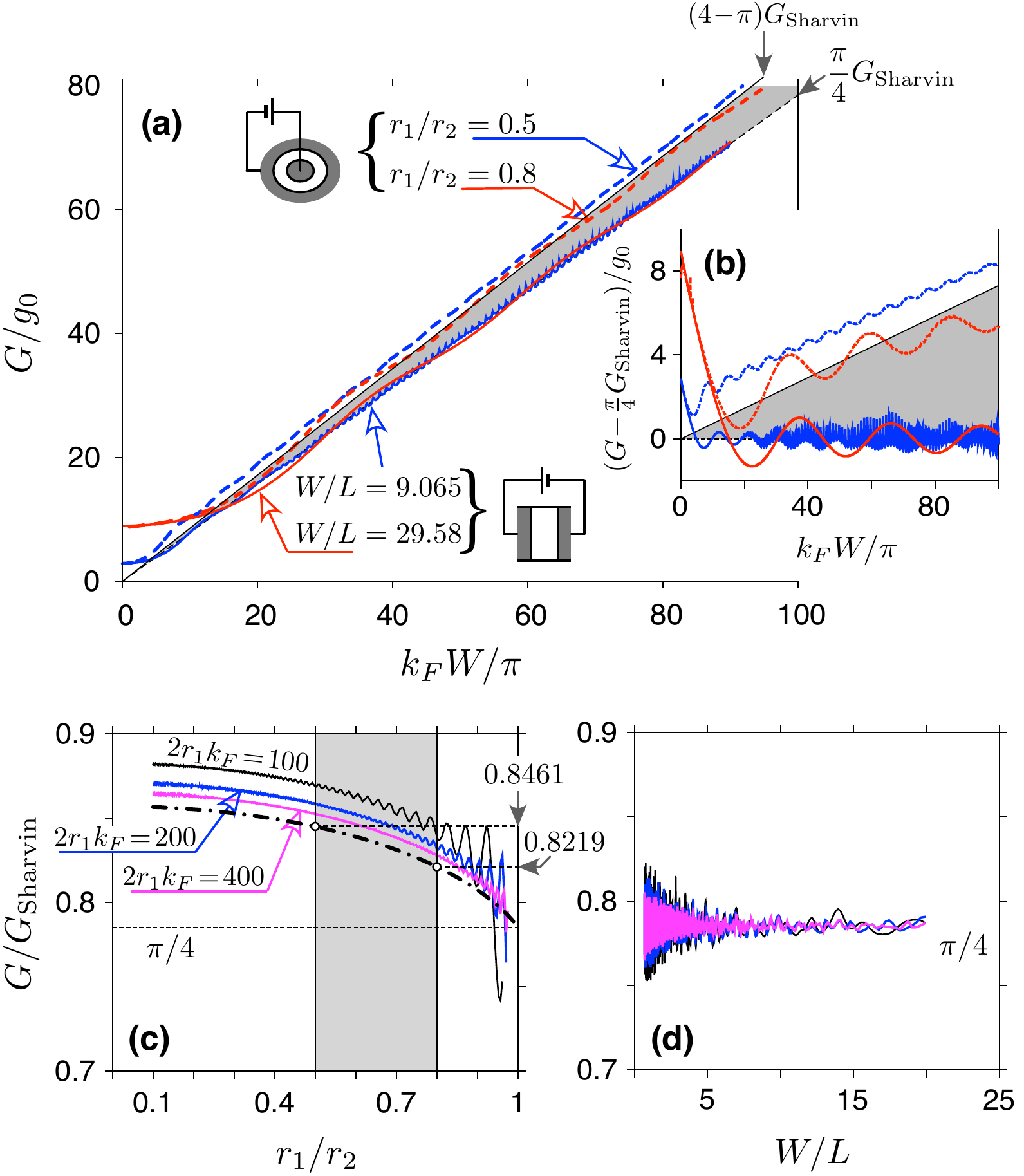}
  \caption{ \label{fig:gsharvd}
    (a) Conductance as a~function of the Fermi momentum for Corbino disks
    with the radii ratios $r_1/r_2=0.5$ and $0.8$ (thick dashed lines) and
    for rectangular graphene samples (thick solid lines) with width-to-length
    ratios adjusted to match the zero-energy conductance, i.e.,
    $W/L=2\pi/\log(r_2/r_1)$, and specified for each line.
    Both types of systems are shown schematically. 
    (b) The excess conductance above the sub-Sharvin value
    $(\pi/4)\,G_{\rm Sharvin}$,
    with $G_{\rm Sharvin}=g_0k_F{}W/\pi$ (for the disks, we set $W=2\pi{}r_1$).
    The conductance quantum is $g_0=4e^2/h$. 
    (c) The conductance reduction as a~function of the radii ratio for disks
    at fixed values of $G_{\rm Sharvin}/g_0=2r_1{}k_F=100$, $200$, and $400$
    (specified for each solid line).
    Dash-dotted line marks the approximating formula given by Eq.\
    (\ref{ggdiskicoh});
    the numerical values for $r_1/r_2=0.5$ and $0.8$ are indicated with open
    symbols and specified up to four decimal places. 
    (d) The conductance reduction as a~function of the aspect ratio for
    rectangles with same values of $G_{\rm Sharvin}/g_0=k_FW/\pi$ as in (c).
    Thin dashed lines in (a)--(d) mark the sub-Sharvin value
    $(\pi/4)\,G_{\rm Sharvin}$. Thin solid lines in (a), (b) depict the asymptotic
    conductance for $r_1/r_2\rightarrow{}0$, being equal to
    $(4-\pi)\,G_{\rm Sharvin}$. 
  }
\end{figure}

\begin{figure}[!t]
  \includegraphics[width=\linewidth]{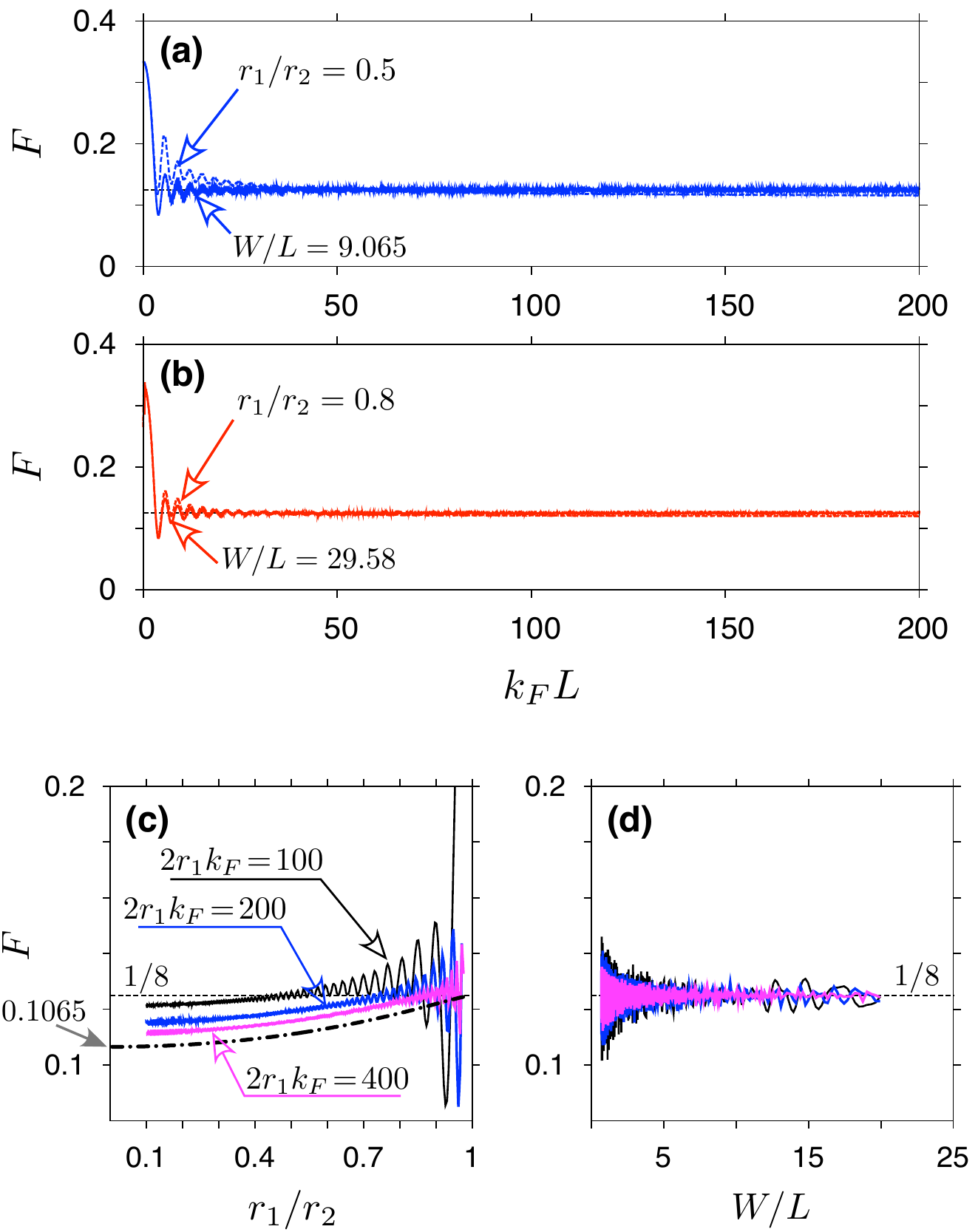}
  \caption{ \label{fig:fafarvd}
    (a,b) Fano factor as a~function of the Fermi momentum for same systems
    as in Figs.\ \ref{fig:gsharvd}(a), \ref{fig:gsharvd}(b).
    The radii ratio (or aspect ratio) is specified for each line.
    The sample length for disks is defined as $L=r_2-r_1$.
    (c) Fano factor as a~function of the radii ratio for disks and
    (d) as a~function of the radii ratio for rectangles with same values
    of $G_{\rm Sharvin}/g_0$ as used in Figs.\ \ref{fig:gsharvd}(c) and
    \ref{fig:gsharvd}(d).
    Dash-dotted line in (c) marks the approximating formula given by Eq.\
    (\ref{ffdiskicoh}); the limiting value for $r_1/r_2\rightarrow{}0$
    is specified up to four decimal places. Thin dashed lines in (a)--(d)
    mark the sub-Sharvin value of $F=1/8$. 
  }
\end{figure}

\begin{figure}[!t]
  \includegraphics[width=0.9\linewidth]{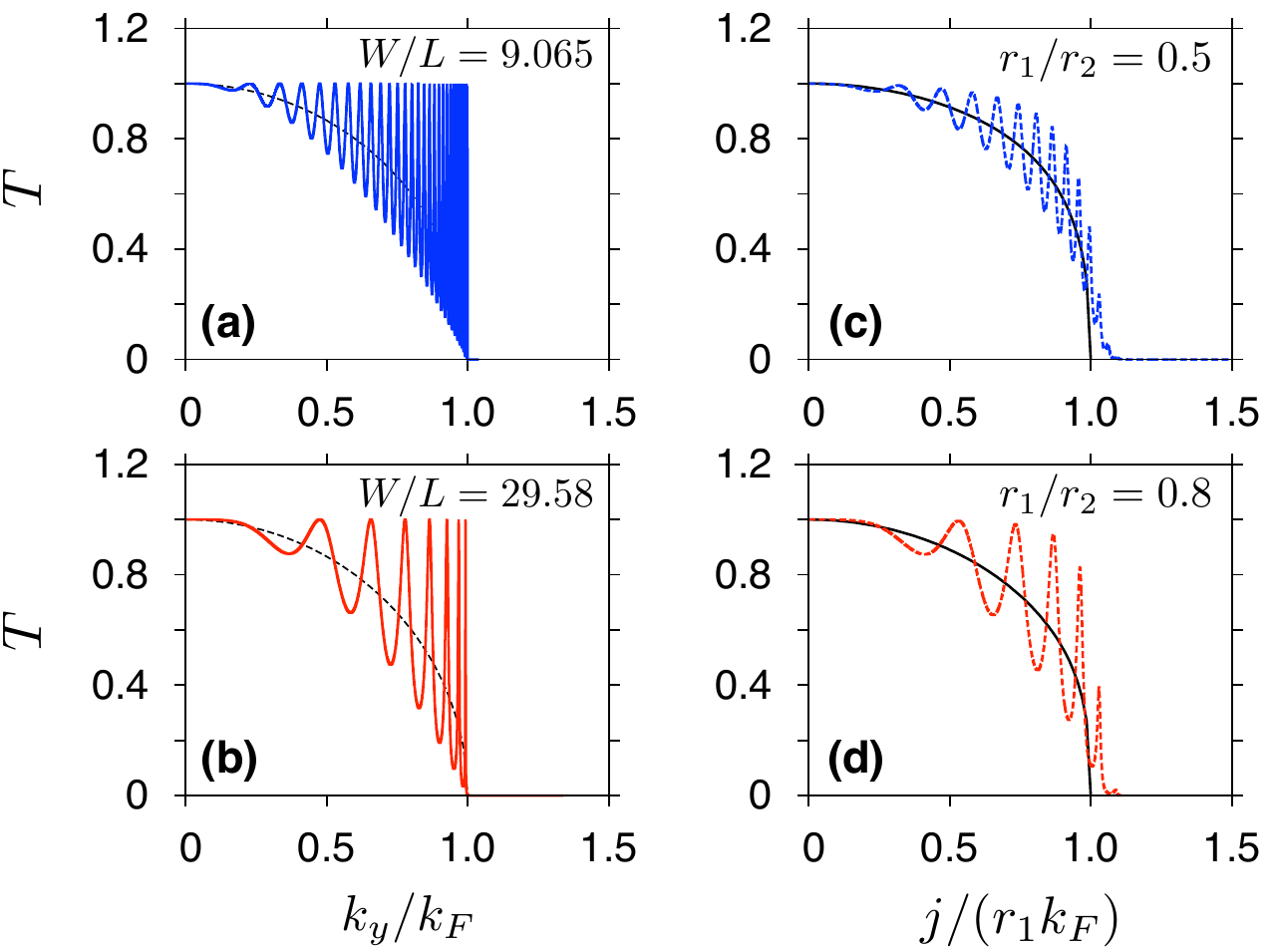}
  \caption{ \label{fig:ttkyrvd}
    Transmission probability as a~function of the transverse momentum
    for rectangular samples 
    (a,b) or the total angular momentum for disks (c,d).
    The Fermi momentum is fixed such that $k_FL=r_1k_F=100$ (a,c)
    or $k_FL=(r_2\!-\!r_1)k_F=25$ (b,d). 
    Remaining system parameters (specified at each panel) are same as
    in Figs.\ \ref{fig:fafarvd}(a) and \ref{fig:fafarvd}(b).
    Thin dashed lines in (a) and (b) mark Eq.\ (\ref{ttrecappr}); 
    solid lines in (c) and (d) mark Eq.\ (\ref{ttdiskicoh}). 
  }
\end{figure}

\subsection{Conductance and Fano factor}
Numerical values of the conductance and the Fano factor, obtained for
the two systems for which exact expressions for transmission probabilities
$T_n$ (or $T_j$) are available, are presented in Figs.\ \ref{fig:gsharvd} and
\ref{fig:fafarvd} \cite{kr1maxfoo}.
For rectangular samples with infinite-mass confinement, 
we simply took the limit of $N\rightarrow{}\infty$ in Eq.\ (\ref{gfland})
numerically. For disks, $T_n$-s in Eq.\ (\ref{gfland}) are replaced with 
$T_j$-s gives by Eqs.\ (\ref{tjphi}) and (\ref{ddnupm}), and the summations
are performed for $-\infty{}<j<\infty$, with $j$ being half-odd integer.

In Fig.\ \ref{fig:gsharvd}(a), the conductance spectra for selected systems
are compared with the asymptotic formulas given in Eqs.\ (\ref{gfasym1})
and (\ref{gfasym0}). Since the limit of an infinite rectangular barrier is
considered, there are only two dimensionless parameters relevant: the radii
(or aspect) ratio $r_1/r_2$ (or $W/L$) and the expected (dimensionless) Sharvin
conductance $G_{\rm Sharvin}/g_0=2k_Fr_1$ (or $\,=k_FW/\pi$). In order to
compare the results for different geometries, we further adjust the aspect
ratio for rectangles such that the zero-energy conductance $G=(4/\pi)g_0W/L$
is the same as for a~given disk, namely: $W/L=2\pi/\log(r_2/r_1)$, with
$r_2/r_1=2$ and $1.25$. 
The last relation can be easily derived by taking the zero-energy limit in
Eq.\ (\ref{tjphi}), $T_j(k\rightarrow{}0)=
4/\left[(r_2/r_1)^j+(r_1/r_2)^j\right]^2$, and approximating the sum over $j$
by an integral \cite{zerokfoo}.

It is easy to see that the conductance spectra for rectangular samples
[see thick solid lines in Figs.\ \ref{fig:gsharvd}(a) and \ref{fig:gsharvd}(b)]
closely follow the sub-Sharvin formula $(\pi/4)\,G_{\rm Sharvin}$ soon after
the condition $k_F\gtrsim{}L^{-1}$ (i.e., the ballistic transport prevails
over the pseudodiffusive transport) is satisfied. In contrast, the conductance
spectra for disks [see thick dashed lines in Figs.\ \ref{fig:gsharvd}(a)
and \ref{fig:gsharvd}(b)] slowly approach the upper limit of 
$(4-\pi)\,G_{\rm Sharvin}$ corresponding to $r_2\gg{}r_1$. In fact, for the
range of $2r_1k_F\leqslant{}100$ used in Figs.\ \ref{fig:gsharvd}(a) and
\ref{fig:gsharvd}(b), the values of $G<(4-\pi)\,G_{\rm Sharvin}$ can be
noticed only for the case of $r_1/r_2=0.8$.

Results for higher values of $2r_1k_F$ are displayed, versus the radii
ratio, in Fig.\ \ref{fig:gsharvd}(c). It becomes clear now that the values
following from approximating Eq.\ (\ref{ggdiskicoh}) for incoherent
transmission [see black dash-dotted line] are approached by exact numerical
results [solid lines] for any $r_1/r_2$. However, the convergence is
noticeably slower than for rectangular samples, see
Fig.\ \ref{fig:gsharvd}(d). In physical units, the inner disk diameter of
$2r_1=1000\,$nm and the maximal Fermi energy of $|E|=0.3\,$eV (already
reported for some graphene-hBN heterostructures, see Ref.\ \cite{Ter16})
correspond to $2r_1k_F\approx{}522$, allowing us to expect that the values
of $(4-\pi)>G/G_{\rm Sharvin}>\pi/4$ should be observable in graphene disks
with moderate radii ratios $0.5\leqslant{}r_1/r_2\leqslant{}0.8$. 

Values of the shot-noise power for the same systems are presented
in Fig.\ \ref{fig:fafarvd}. This time, we display the Fano factor as 
a~function of $k_FL$ (with $L=r_2-r_1$ for disks) in order to visualize
quasiperiodic oscillations of the Fabry-P\'{e}rot type, which are
well-pronounced for both rectangular and disk-shaped samples with different
aspect (or radii) ratios, see Figs.\ \ref{fig:fafarvd}(a) and
\ref{fig:fafarvd}(b). 
Similarly as for the conductance, the Fano factor for rectangular samples
[solid lines] shows fast convergence, with growing $k_F$, to the sub-Sharvin
value of $F=1/8$.
For disks, the convergence is slower, and the limiting value for
large $k_F$ is significantly lower than $1/8$ [dashed lines].

Again, plotting the Fano factor for several fixed $2r_1k_F$ and varying
$r_1/r_2$ [see Fig.\ \ref{fig:fafarvd}(c)], allows us to notice an apparent
convergence of exact numerical results [solid lines] to predictions
following from Eq.\ (\ref{ffdiskicoh}) [dash-dotted line]. Also,
the above-mentioned convergence is noticeably slower for disks that
for rectangular samples [see Fig.\ \ref{fig:fafarvd}(d)].

For better understanding of the effects described in this section we plot,
in Fig.\ \ref{fig:ttkyrvd}, transmission probabilities given by
Eq.\ (\ref{ttkxky}) for rectangles, or by Eqs.\ (\ref{tjphi}) and
(\ref{ddnupm}) for disks, as functions of $k_y/k_F$ or $j/(r_1k_F)$
(respectively) at a~fixed $k_F$. It worth to point out, that Eqs.\
(\ref{tjphi}) and (\ref{ddnupm}) are valid for any fractional value of $j$;
physically, other than half-odd integer $j$-s may appear in the presence of
magnetic flux piercing the inner electrode, see Ref.\ \cite{Ryc20}. 

Remarkably, transmission probability for evanescent modes, with
$j/r_1>k_F$, decays (with growing $j$) significantly slower for disks
[see Figs.\ \ref{fig:ttkyrvd}(c) and \ref{fig:ttkyrvd}(d)] than for
rectangular samples, for which we immediately have $T\approx{}0$ if
$k_y>k_F$ [see Figs.\ \ref{fig:ttkyrvd}(a) and \ref{fig:ttkyrvd}(b)].
This is the reason, for which charge-transfer characteristics
obtained via Eq.\ (\ref{gfland}) for disks generically show slower
convergence (with growing $k_F$) to the predictions for incoherent
transmission presented in Sec.\ \ref{appcofan} than it can be observed
in corresponding data for rectangles.

\begin{figure}[!b]
  \includegraphics[width=0.7\linewidth]{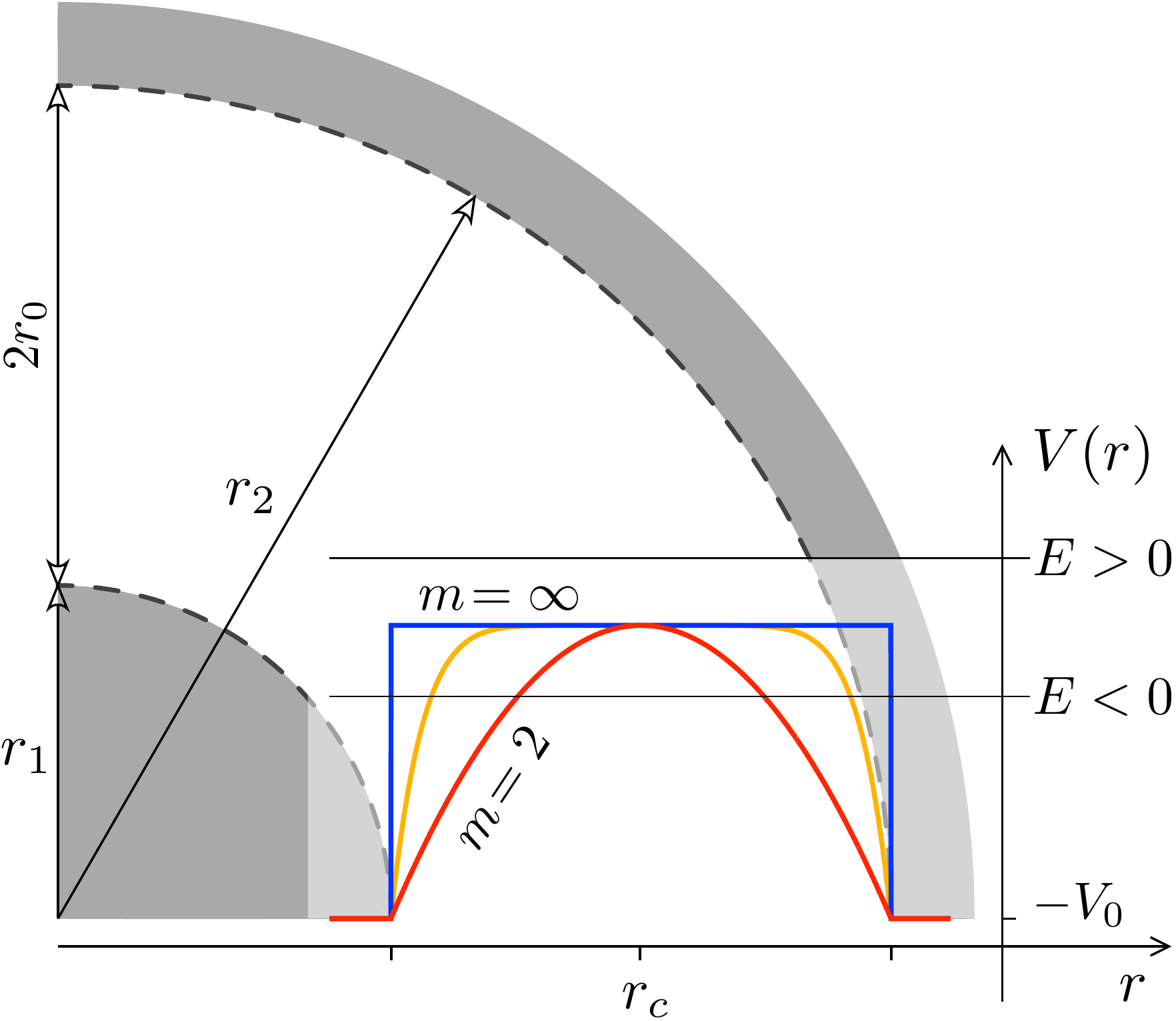}
  \caption{ \label{diskmpot}
    Electrostatic potential profiles given by Eq.\ (\ref{v0mpot}) with
    $m=2,8$ and $m=\infty$ (i.e., the rectangular barrier). 
    The Fermi energy $E$ is defined with respect to the top of a~barrier.
    $E>0$ corresponds to unipolar n-n-n doping in the device;
    for $E<0$, circular n-p-n structure is formed.
    Arcs with the radii $r_1$ and $r_2$ (dashed lines) mark the interfaces
    between the disk area [$r_1<r<r_2$; white area] and contact regions
    [$r<r_1$ or $r>r_2$; shaded areas].
  }
\end{figure}

\section{Smooth potential barriers}
\label{smoopots}

\subsection{Mode-matching for smooth potentials}
For the sake of completeness, we also revisit, in this section, the
effects of smooth potential barriers, earlier considered for the
rectangular geometry \cite{Ryc21b}. For the Corbino disk, key steps of
the reasoning remain similar as presented in Sec.\ \ref{exadisk}A.
However, the electrostatic potential energy in Eqs.\ (\ref{phapri}) and
(\ref{phbpri}) is now replaced by
\begin{equation}
  \label{v0mpot}
  V(r) = -V_0\times
  \begin{cases}
    \,\left|(r-r_c)/r_0\right|^m  &  \text{if }\ |r-r_c| \leqslant r_0, \\
    \,1  &  \text{if }\ |r-r_c| > r_0, 
  \end{cases}
\end{equation}
where $r_c=(r_1+r_2)/2$ and $r_0=(r_2-r_1)/2$. 
Changing the value of $m$ tunes the potential from parabolic shape ($m=2$)
to rectangular shape ($m\rightarrow\infty$), see Fig.\ \ref{diskmpot}. 

Since $V(r)=-V_0$ for $r<r_1$ and $r>r_2$, solutions for the leads given
by Eq.\ (\ref{chijleads}) remain unchanged. This time, we do not take the
limit of $V_0\rightarrow\infty$; instead, $V_0=t_0/2=1.35\,$eV (being close
the values appearing in first-principle calculations \cite{Gio08,Cus17})
is considered in subsequent numerical examples.
In turn, wavefunctions in the leads can now be written as follows
\begin{align}
  \chi_j^{(1)} &=
  \chi_j^{\rm in} + r_j\chi_j^{\rm out},
    \  &r<r_1,
  \label{chij1hank} \\
  \chi_j^{(2)} &= t_j\chi_j^{\rm in}, 
    \  &r>r_2, \label{chij2hank} 
\end{align}
where $\chi_j^{\rm in}$, $\chi_j^{\rm out}$ are given by
Eq.\ (\ref{chijleads}) with $k\equiv{}k_0=|E+V_0|/(\hbar{}v_F)$
\cite{hankfoo} and
$r_j$ ($t_j$) denotes the reflection (transmission) amplitudes. 

In the disk area, $r_1<r<r_2$, $V(r)$ given by Eq.\ (\ref{v0mpot}) is no
longer piecewise-constant, and Eqs.\ (\ref{phapri}) and
(\ref{phbpri}) need to be integrated numerically for
$j=\pm{}\frac{1}{2},\pm\frac{3}{2},\dots$ \cite{simfoo}. 
The resulting wavefunction takes a~form
\begin{equation}
\label{chijdisknum}
  \chi_j^{(d)}=A\chi_j^{\rm I}+B\chi_j^{\rm II}, 
\end{equation}
where $\chi_j^{\rm I}$, $\chi_j^{\rm II}$ denote the two linearly independent
solutions, 
which we obtained numerically by solving the relevant equations assuming
two different initial conditions
$\left.\chi_j^{{\rm I}, {\rm II}}\right|_{r=r_1}=(1,\pm{}1)^T$, and $A$, $B$ are
arbitrary complex coefficients. 

The matching conditions for $r=r_1$ and $r=r_2$ brought us to the linear
system of equations for $A$, $B$, $r_j$, and $t_j$,
\begin{widetext}
\begin{equation}
  \label{lsysABrt}
  \left[
    \begin{matrix}
      \chi_{j,a}^{\rm out}(r_1) & -\chi_{j,a}^{\rm I}(r_1)
      & -\chi_{j,a}^{\rm II}(r_1) & 0 \\
      \chi_{j,b}^{\rm out}(r_1) & -\chi_{j,b}^{\rm I}(r_1)
      & -\chi_{j,b}^{\rm II}(r_1) & 0 \\
      0 &  -\chi_a^{\rm I}(r_2)  & -\chi_{j,a}^{\rm II}(r_2)
      & \chi_{j,a}^{\rm in}(r_2) \\
      0 &  -\chi_b^{\rm I}(r_2)  & -\chi_{j,b}^{\rm II}(r_2)
      & \chi_{j,b}^{\rm in}(r_2) \\
    \end{matrix}
  \right]
  \left[
    \begin{matrix}
      r_j \\ A \\ B \\ t_j \\
    \end{matrix}
  \right]
  =
  \left[
    \begin{matrix}
      -\chi_{j,a}^{\rm in}(r_1) \\
      -\chi_{j,b}^{\rm in}(r_1) \\
      0 \\
      0 \\
    \end{matrix}
  \right], 
\end{equation}
\end{widetext}
where we have explicitly written the spinor components of relevant
wavefunctions appearing on right-hand sides of Eqs.\ (\ref{chij1hank}),
(\ref{chij2hank}), and (\ref{chijdisknum}).
Since linear systems of the form given by Eq.\ (\ref{lsysABrt}) for
different $j$-s are decoupled, numerous software packages can be employ
find their solutions up to a~machine precision \cite{zgesv99}.

\subsection{Sharvin conductance}
Before presenting the numerical results obtained by finding the
transmission probabilities $T_j=|t_j|^2$ from Eq.\ (\ref{lsysABrt}), 
we first comment how to define the Sharvin conductance for a~disk-shaped
system subjected to the electrostatic potential $V(r)$ given by
Eq.\ (\ref{v0mpot}). In such a~case, we simply look for the minimal number
of propagating modes in the disk area, writing
\begin{equation}
  \label{gsharm}
  G_{\rm Sharvin}^{(m)}=2g_0\min_{r_1\leqslant{}r\leqslant{}r_2}\left[\,rk_F(r)\,\right], 
\end{equation}
where 
\begin{equation}
  \label{kfrdef}
  k_F(r)=\frac{|E-V(r)|}{\hbar{}v_F},
\end{equation}
and the index $m$ in Eq.\ (\ref{gsharm}) is the exponent defining $V(r)$.
The Fermi wavenumber $k_F(r)$ is now position-dependent; hereinafter, using
a~simplified symbol ($k_F$) we always refer to $k_F(r_c)=|E|/(\hbar{}v_F)$. 

For general values of $r_1$, $r_2$, and $m$, the minimum in
Eq.\ (\ref{gsharm}) needs to be determined numerically. In the limit of
$m\rightarrow{}\infty$, the minimum corresponds to $r_{\rm min}=r_1$ and we
get
\begin{equation}
  \label{gsharinf}
  G_{\rm Sharvin}^{(\infty)}=2g_0r_1k_F, 
\end{equation}
restoring the formula for a~rectangular barrier introduced in
Eq.\ (\ref{ggdiskicoh}). 
For $m=2$ (parabolic barrier) and $0\leqslant{}E\ll{}V_0$, one can
easily find that $r_{\rm min}\approx{}r_c$ \cite{rmin2foo}, and that
%\begin{equation}
%  \label{rminexp}
%  r_{\rm min} = {}r_c - r_0\left[\frac{r_0E}{2r_cV_0} +
%  {\cal O}\left(\left(\frac{r_0E}{r_cV_0}\right)^2\right)\right], 
%\end{equation}
the following approximation
\begin{equation}
  \label{gshar2}
  G_{\rm Sharvin}^{(2)}\approx{}2g_0r_ck_F=(r_c/r_1)\,G_{\rm Sharvin}^{(\infty)}
\end{equation}
should be sufficient for typical, experimentally-accessible, values of
$r_1/r_2$ and $E$.

\begin{figure}[!t]
  \includegraphics[width=0.9\linewidth]{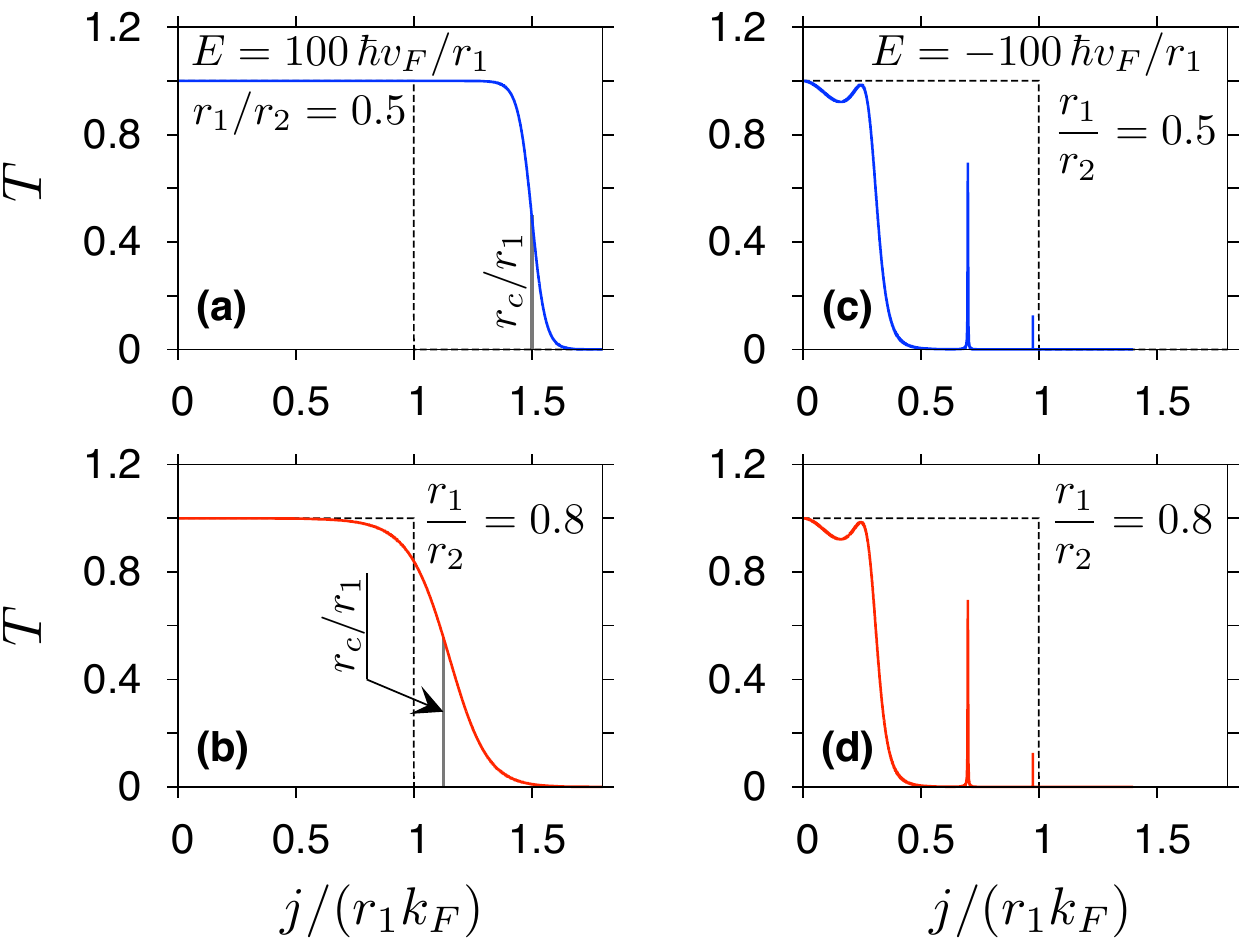}
  \caption{ \label{fig:ttkydim2}
    Transmission probability versus angular momentum for disks with
    parabolic potential barrier given by Eq.\ (\ref{v0mpot}) with $m=2$.
    The scattering energy is fixed at $E=100\,\hbar{}v_F/r_1$ (a,b) or
    $E=-100\,\hbar{}v_F/r_1$ (c,d). The radii ratio is specified at each panel. 
    The remaining parameters are $r_1=800\,$nm and $V_0=1.35\,$eV.
    Dashed lines depict the step functions
    $T_j=\Theta\left(k_F-|j|/r_1\right)$, with $k_F\equiv{}k_F(r_c)=
    |E|/(\hbar{}v_F)$ [see Eq.\ (\ref{kfrdef})].
    Vertical lines in (a,b) shows the ratio $r_c/r_1=1.5$ (a) or $1.125$ (b).
  }
\end{figure}

\begin{figure}[!t]
  \includegraphics[width=0.9\linewidth]{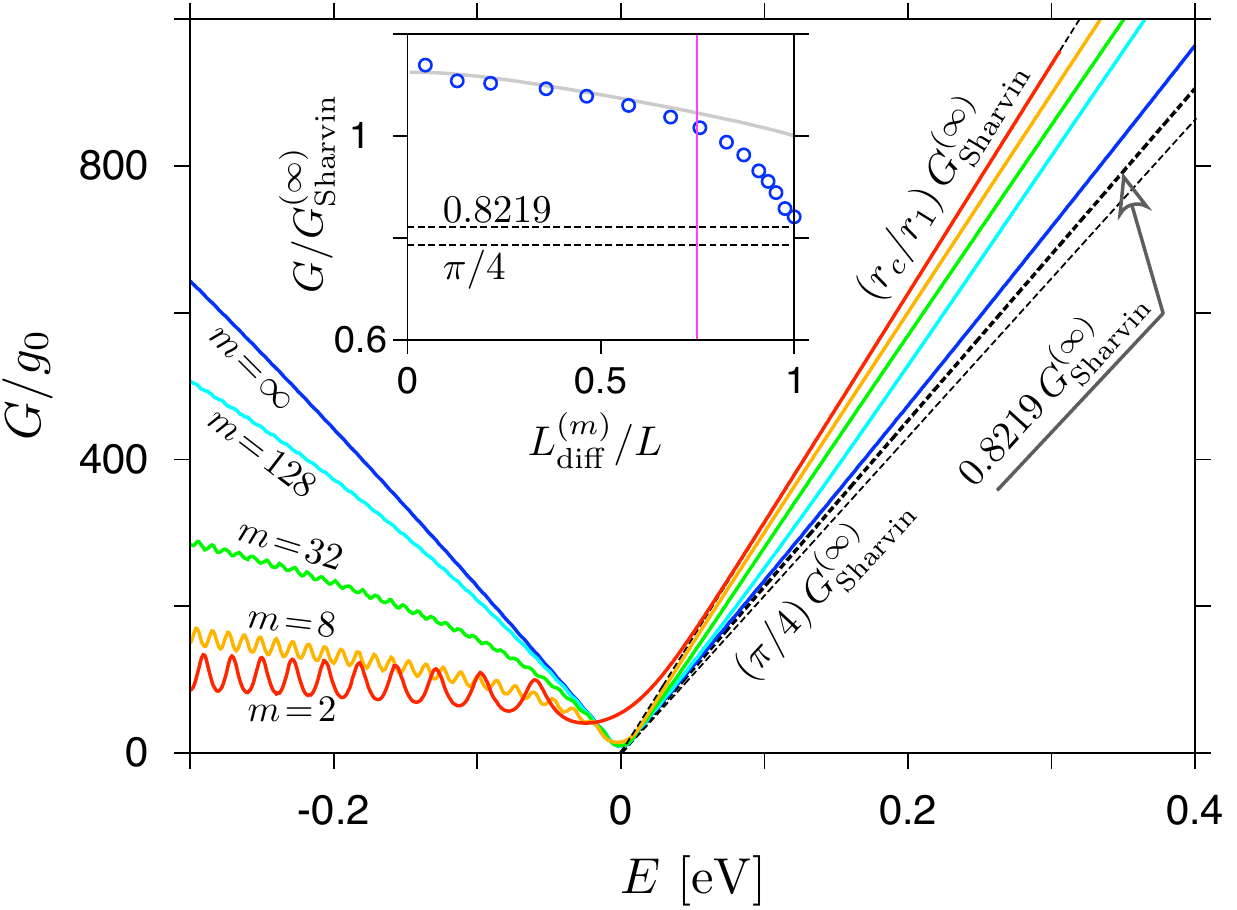}
  \caption{ \label{fig:gr800vsen}
    Main: Conductance as a~function of the Fermi energy for the arrangement
    depicted in Fig.\ \ref{diskmpot}. The disk radii are $r_1=800\,$nm
    and $r_2=1000\,$nm, the barrier height is $V_0=t_0/2=1.35\,$eV.
    The exponent $m$ in Eq.\ (\ref{v0mpot}) is specified for each dataset
    (solid lines).
    Dashed lines depict the approximating Eq.\ (\ref{gshar2}) for $m=2$
    with $G_{\rm Sharvin}^{(\infty)}=2g_0r_1|E|/(\hbar{}v_F)$, the asymptotic
    formula in Eq.\ (\ref{gfasym1}) for $r_1/r_2\rightarrow{}1$ [thin dashed
    lines], and the expression following from Eq.\ (\ref{ggdiskicoh})
    for $r_1/r_2=0.8$ [thick dashed line]. 
    Inset: Conductance at $E=100\,\hbar{}v_F/r_1\approx{}72\,$meV as
    a~function of $L_{\rm diff}^{(m)}/L$, with $L\equiv{}r_2-r_1$, see Eq.\
    (\ref{ldiffm}). 
    Grey solid line depicts $G_{\rm Sharvin}$ obtained from Eq.\ (\ref{gsharm})
    by numerical minimization. 
    Vertical line marks a~bound on the right-hand side of Eq.\
    (\ref{ldiffbound}). 
  }
\end{figure}

\begin{figure}[!t]
  \includegraphics[width=0.9\linewidth]{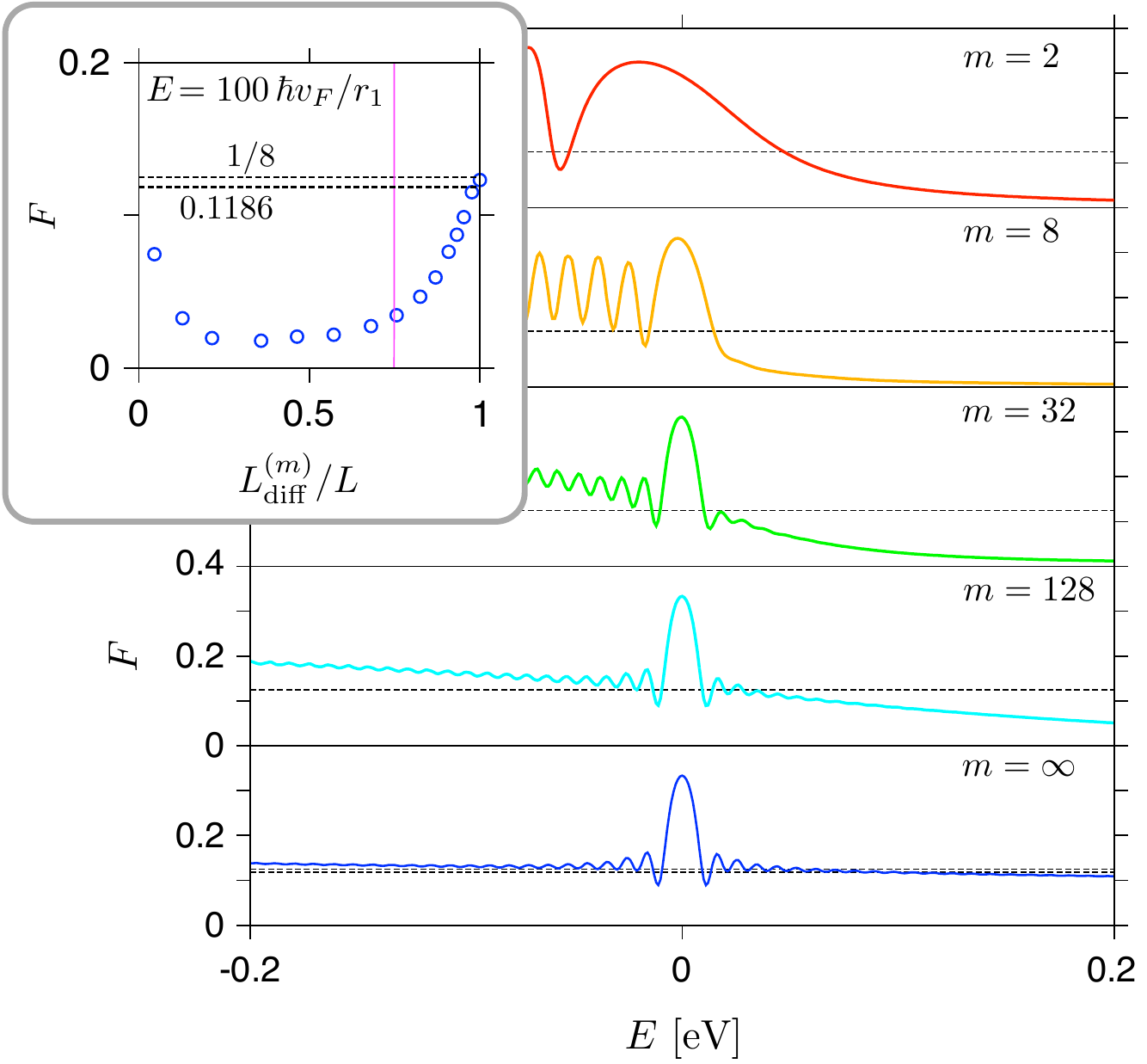}
  \caption{ \label{fig:ffr800vsen}
    Fano factor as a function of the Fermi energy for the same
    system parameters as in Fig.\ \ref{fig:gr800vsen} (solid lines).
    The exponent $m$ in Eq.\ (\ref{v0mpot}) is varied between the panels.
    Thin dashed line at each panel depicts the sub-Sharvin value of $F=1/8$,
    see Eq.\ (\ref{gfasym1}). Thick dashed line in the bottom panel
    marks the value obtained from Eq.\ (\ref{ffdiskicoh})
    for $r_1/r_2=0.8$. 
    Inset: Fano factor at $E=100\,\hbar{}v_F/r_1\approx{}72\,$meV as
    a~function of $L_{\rm diff}^{(m)}$, see Eq.\ (\ref{ldiffm}).
    Vertical line marks a~bound on the right-hand side of Eq.\
    (\ref{ldiffbound}). 
  }
\end{figure}

\subsection{Numerical results}
In Fig.\ \ref{fig:ttkydim2}, we display transmission probabilities for
systems with $r_1=800\,$nm and two different values of $r_2=1600\,$nm and
$1000\,$nm, corresponding to $r_1/r_2=0.5$ and $0.8$, as functions of $j$. 
(Fractional, i.e., other than half-odd integer values of $j$ in Eq.\
(\ref{lsysABrt}) have no physical meaning here and are considered for
plotting purposes only.)
The parabolic barrier ($m=2$) is considered in all cases.
The Fermi energy is fixed at $E=100\,\hbar{}v_F/r_1$ or at
$E=-100\,\hbar{}v_F/r_1$ , corresponding to $E\approx\pm{}72\,$meV
in the physical units.

For $E>0$ [see Figs.\ \ref{fig:ttkydim2}(a) and \ref{fig:ttkydim2}(b)],
transmission essentially shows a~familiar switching behavior \cite{Kem35},
with $T\approx{}1$ for $j/r_c<k_F$ and $T\approx{}0$ for $j/r_c>k_F$.
Notice that, comparing to the case of a~rectangular barrier discussed
in Sec.\ \ref{exadisk}B [see Figs.\ \ref{fig:ttkyrvd}(c) and
\ref{fig:ttkyrvd}(d)], $r_c$ now plays a~role similar to $r_1$, 
coinciding with a~prediction given in Eq.\ (\ref{gshar2}). 
For $E<0$, see Figs.\ \ref{fig:ttkydim2}(c) and \ref{fig:ttkydim2}(d),
the presence of two circular p-n junctions [positioned at $r=r_{\star}$ such
that $V(r_{\star})=E$, see Fig.\ \ref{diskmpot}] significantly reduces the
transmission for almost any $j$.

Numerical results for the conductance and Fano factor, obtained by summing
over the modes [see Eq.\ (\ref{gfland})] with half-odd integer $j$, are
presented in Figs.\ \ref{fig:gr800vsen} and \ref{fig:ffr800vsen}.
We consider the radii ratio of $r_1/r_2=0.8$ now, in order to find out
whether (or not) the modifications to sub-Sharvin charge-transfer
characteristics, described in Secs.\ \ref{appcofan} and \ref{exadisk},
are still significant in such a~relatively thin disk subjected to smooth
potential barrier of a~finite height.

Substituting $a=r_1/r_2=0.8$ into Eqs.\ (\ref{ggdiskicoh}) and
(\ref{ffdiskicoh}) we obtain, respectively,
\begin{equation}
  \label{gficoh:a08}
  G/G_{\rm Sharvin}^{(\infty)}\approx{}0.8219
  \ \ \ \ \ \ \text{and}\ \ \ \ \ \ 
  F\approx{}0.1186.
\end{equation}
Both predictions differ by about $5\%$ from the asymptotic values
given in Eq.\ (\ref{gfasym1}) and corresponding to $V_0\rightarrow{}\infty$,
$m\rightarrow{}\infty$, and $r_1/r_2\rightarrow{}1$; the conductance is  
expected to be elevated, whereas noise is expected to be suppressed
for $r_1/r_2<1$.
Although Eqs.\ (\ref{ggdiskicoh}) and (\ref{ffdiskicoh}) are proposed,
still as approximations, for a~perfectly rectangular barrier of an infinite 
height, numerical results for a~{\em finite} $V_0=1.35\,$eV, and $m=\infty$
[see blue solid lines in Figs.\ \ref{fig:gr800vsen} and
\ref{fig:ffr800vsen}], are relatively close to the predictions given in
Eq.\ (\ref{gficoh:a08}).
This observation applies particularly for
$E\sim{}0.1\,$eV, i.e., for $E_{\rm diff}\ll{}E\ll{}V_0$, where
\begin{equation}
\label{efdifdef}
  E_{\rm diff}=\frac{\hbar{}v_F}{L}\approx{}3\,\text{meV}\ \ \
  \text{for}\ \ \ L\equiv{}r_2\!-\!r_1=200\,\text{nm}
\end{equation}
denotes the energy above which Sharvin conductance overrules the
pseudodiffusive conductance, see Ref.\ \cite{Ryc21b}. 
Remarkably, the conductance in such a~range is definitely closer to the
value given in Eq.\ (\ref{gficoh:a08}) than to Eq.\ (\ref{gfasym1}).

For the Fano factor (see Fig.\ \ref{fig:ffr800vsen}), the situation is less
clear due to oscillations of the Fabry-P\'{e}rrot type with an amplitude
(albeit being reduced in comparison to the rectangular
geometry, see Ref.\ \cite{Ryc21b}) exceeding the distance between the
predictions given in Eqs.\ (\ref{gfasym1}) and (\ref{gficoh:a08}).
Therefore, when looking for a~finite radii-ratio
effects on the shot-noise power, one should rather focus on the $r_2\gg{}r_1$
range, where the predicted suppression of $F$, comparing 
Eqs.\ (\ref{gfasym1}) and (\ref{gfasym0}), is close to $15\%$. 

A~striking feature of the data presented in Figs.\ \ref{fig:gr800vsen}
and \ref{fig:ffr800vsen} is a~systematic evolution, for $E\gg{}E_{\rm diff}$,
towards the values of $G$ given by Eq.\ (\ref{gsharm}) and $F\approx{}0$, 
when decreasing the value of $m$, i.e., tuning the potential barrier from
rectangular ($m=\infty$) towards parabolic ($m=2$) shape. (Notice that
red solid line in Fig.\ \ref{fig:gr800vsen}, representing the results of our
numerical mode-matching for $m=2$, precisely covers dashed line marking the
approximating Eq.\ (\ref{gshar2}) for almost the entire range of $E>0$
presented in the plot.)

For $E<0$, two circular p-n junctions reduce the transmission for any finite
$m$, resulting in the suppressed conductance (see Fig.\ \ref{fig:gr800vsen})
and the enhanced Fano factor (see Fig.\ \ref{fig:ffr800vsen}), with
strong oscillations due to quasibound states \cite{Sil07}. 

Since the Sharvin conductance for a~disk setup ($G_{\rm Sharvin}^{(m)}$) is
$m$-dependent, see Eqs.\ (\ref{gsharm}) and (\ref{kfrdef}), it is worth to
introduce the effective sample length evolving with $m$, such that
$L_{\rm diff}=r_2-r_1$ for $m=\infty$ (rectangular barrier),
and $L_{\rm diff}\ll{}r_2-r_1$ for $m=2$ (parabolic barrier).
(In the latter case, a~narrow weakly-doped ring is placed near the distance
of $r=r_c$ from the disk center, allowing one to understand why
the approximation given in Eq.\ (\ref{gshar2}) works well for
$E\gg{}E_{\rm diff}$.) The effective length can be defined via $E_{\rm diff}$
(\ref{efdifdef}) by imposing $V(\pm{}L_{\rm diff}/2)=-E_{\rm diff}$, leading to
\begin{equation}
  \label{ldiffm}
  {L_{\rm diff}^{(m)}} = 2r_0\left(\frac{\hbar{}v_F}{2r_0V_0}\right)^{1/m}.  
\end{equation}
One easily finds that the above reduces to
$L_{\rm diff}^{(\infty)}=2r_0=r_2-r_1\equiv{}L$ for a~rectangular barrier; also,
we have $L_{\rm diff}^{(2)}=L\sqrt{E_{\rm diff}/V_0}\ll{}L$.
Subsequently, characteristic length scale of a~potential jump
$\Delta{}r=(L-L_{\rm diff})/2$, can be compared with the Fermi wavelength
$\lambda_F=2\pi/k_F(r_c)=h{}v_F/|E|$, allowing to expect that for
$\lambda_F/2\lesssim{}\Delta{}r$ the barrier cannot longer be regarded as
rectangular. The last condition can be rewritten as
\begin{equation}
  \label{ldiffbound}
  \frac{L_{\rm diff}}{L}\lesssim{} 1-\frac{\pi\hbar{}v_F}{r_0|E|},  
\end{equation}
giving $L_{\rm diff}/L\lesssim{}0.7487$ for $E=100\,\hbar{}v_F/r_1$ and
the remaining parameters as used in Figs.\ \ref{fig:gr800vsen} and
\ref{fig:ffr800vsen}. 

Inset in Fig.\ \ref{fig:gr800vsen}, where we display the conductance
for the Fermi energy fixed at $E=100\,\hbar{}v_F/r_1$ (such that
$G_{\rm Sharvin}^{(\infty)}=200\,g_0$) as a~function of $L_{\rm diff}$, 
unveils a~clear switching behavior ruled by the inequality in Eq.\
(\ref{ldiffbound}): For $L_{\rm diff}$ below the upper bound, the datapoints
(representing the results of numerical mode-matching for selected integer
$m$-s) closely follow $G_{\rm Sharvin}$ obtained by performing the minimization
in Eq.\ (\ref{gsharm}) [grey solid line]. For $L_{\rm diff}$ exceeding the
bound, $G$ shows a~fast convergence to the value expected for a~rectangular
barrier ($m=\infty$) and given explicitly in
Eq.\ (\ref{gficoh:a08}).

Similarly, the Fano factor for $E=100\,\hbar{}v_F/r_1$ (see inset in Fig.\
\ref{fig:ffr800vsen}) remains close to $F\approx{}0$ for $L_{\rm diff}$ below
the bound in Eq.\ (\ref{ldiffbound}); above the bound, $F$ converges to
the limiting value lying between the prediction in Eq.\ (\ref{gficoh:a08})
and $F=1/8$ [see Eq.\ (\ref{gfasym1}) for $r_1/r_2\rightarrow{}1$]. 
Elevated values of $F$ for $L_{\rm diff}/L\lesssim{}0.1$ signal a~significant
role of the evanescent modes (with $0<T_j\ll{}1$), which may affect
the noise much stronger than the conductance; see Eq.\ (\ref{gfland}).

\section{A~section of the disk with infinite-mass boundaries}
\label{asecdisk}

\begin{figure}[!h]
  \includegraphics[width=0.35\linewidth]{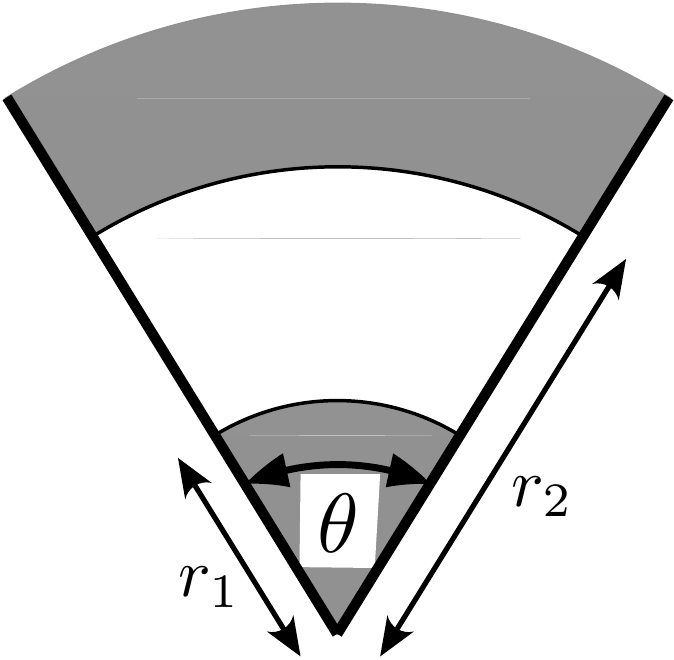}
  \caption{ \label{fig:setupthx}
    A~section of the Corbino disk in graphene (white area) attached
    to heavily-doped leads (shadow areas) and bounded with infinite-mass
    confinement (tick lines). The opening angle $\theta=\pi/3$ and the
    radii ratio $r_1/r_2=0.5$ are set for an illustration. 
  }
\end{figure}

Whole the derivation presented in Sec.\ \ref{appcofan}, in particular,
the approximating formulas for the conductance and Fano factor given
in Eqs.\ (\ref{ggdiskicoh}) and (\ref{ffdiskicoh}), can be easily extended
onto a~section of the disk bounded with infinite-mass confinement, as shown
in Fig.\ \ref{fig:setupthx}. 
Since the absolute value of transverse momentum $|k_y|$ does not
change after a~collision with the boundary \cite{Ber87}, key results
following from the double-contact formula for incoherent transmission,
see Eqs.\ (\ref{tticoh}) and (\ref{tt2icoh}), remain unaltered.
For the reasons which become clear later in this Section, one only needs
to replace the value of Sharvin conductance in Eq.\ (\ref{ggdiskicoh}) by
\begin{equation}
  \label{gsharthx}
  %G_{\rm Sharvin} = \left({\theta}/{\pi}\right)\,k_F{}r_1,
  G_{\rm Sharvin} = \frac{\theta}{\pi}\,g_0r_1{}k_F, 
\end{equation}
with the opening angle $\theta<2\pi$, the inner radii $r_1$, and
the Fermi momentum $k_F$. (Notice that we limit our considerations to the
rectangular potential barrier, leading uniquely-defined $k_F$ for the entire
sample area). For the Fano factor, Eq.\ (\ref{ffdiskicoh}) holds true
for a~bounded disk section as well. 

Details of the mode-matching for coherent scattering of Dirac fermions
in the system of Fig.\ \ref{fig:setupthx} are presented in Ref.\ \cite{Rut14}. 
Here we only recall the main formulas allowing one to determine transmission
eigenvalues for a~discrete set of $\theta$-s and arbitrary dimensionless
parameters $k_Fr_1$, $r_1/r_2$. 

Having in mind the solution for the full disk presented in Sec.\
\ref{exadisk}, we now introduce the sample edges via infinite-mass boundary
conditions. After Berry and Mondragon \cite{Ber87}, we impose that
the angular current vanishes at the sample edges, namely
\begin{equation}
  \left(\mbox{\boldmath$j$}\right)_n=
  \hat{\mbox{\boldmath$n$}}\cdot{}\left[
    \Psi^\dagger\left(\hat{x}\sigma_x+\hat{y}\sigma_y\right) \Psi
  \right]=0, 
\end{equation}
where $\hat{\mbox{\boldmath$n$}}=(\cos\alpha,\sin\alpha)$ is the unit vector
normal to the boundary, the spinor wavefunction $\Psi=(\Psi_a,\Psi_b)^T$, and
the remaining symbols are same as in Eq.\ (\ref{direqvr}).
This leads to \cite{vallfoo}
\begin{equation}
  \label{psapsb}
  \Phi_b/\Phi_a = i\exp\left({i\alpha}\right),
\end{equation}
where $\alpha=0$ for one edge (i.e., at $\varphi=\pi/2$) or
$\alpha=\pi+\theta$ for the other (at $\varphi=\theta+\pi/2$).
The solutions, being linear combinations of the form $a_j\Psi_j+b_j\Psi_{-j}$,
with $\Psi_j=e^{j(j-1/2)\varphi}(\chi_a,\chi_b{}e^{i\varphi})^T$ again, can be found
for a~discrete set of opening angles $\theta\equiv\theta_l=\pi/(2l+1)$, with
$l=0,1,2,\dots$. Explicit formulas for wavefunctions are rather lengthy and
omitted here (see {\it Appendix~A} in Ref.\ \cite{Rut14} for details);
instead, we summarize their basic features as follows:
(i) Due to Eq.\ (\ref{psapsb}), the values of $j$ contributing
to charge-transfer characteristics are now restricted to
\begin{equation}
  \label{jquanthx}
  j = \frac{\pi(2n+1)}{2\theta}, \ \ \ \ n=0,1,2,\dots, 
\end{equation}
justifying the prefactor in Eq.\ (\ref{gsharthx}). 
(ii) Assuming the infinite doping in the leads, 
Transmission probabilities can still be calculated from Eqs.\
(\ref{tjphi}) and (\ref{ddnupm}), with the angular-momentum quantization
given by the above.

\begin{figure}[!t]
  \includegraphics[width=0.9\linewidth]{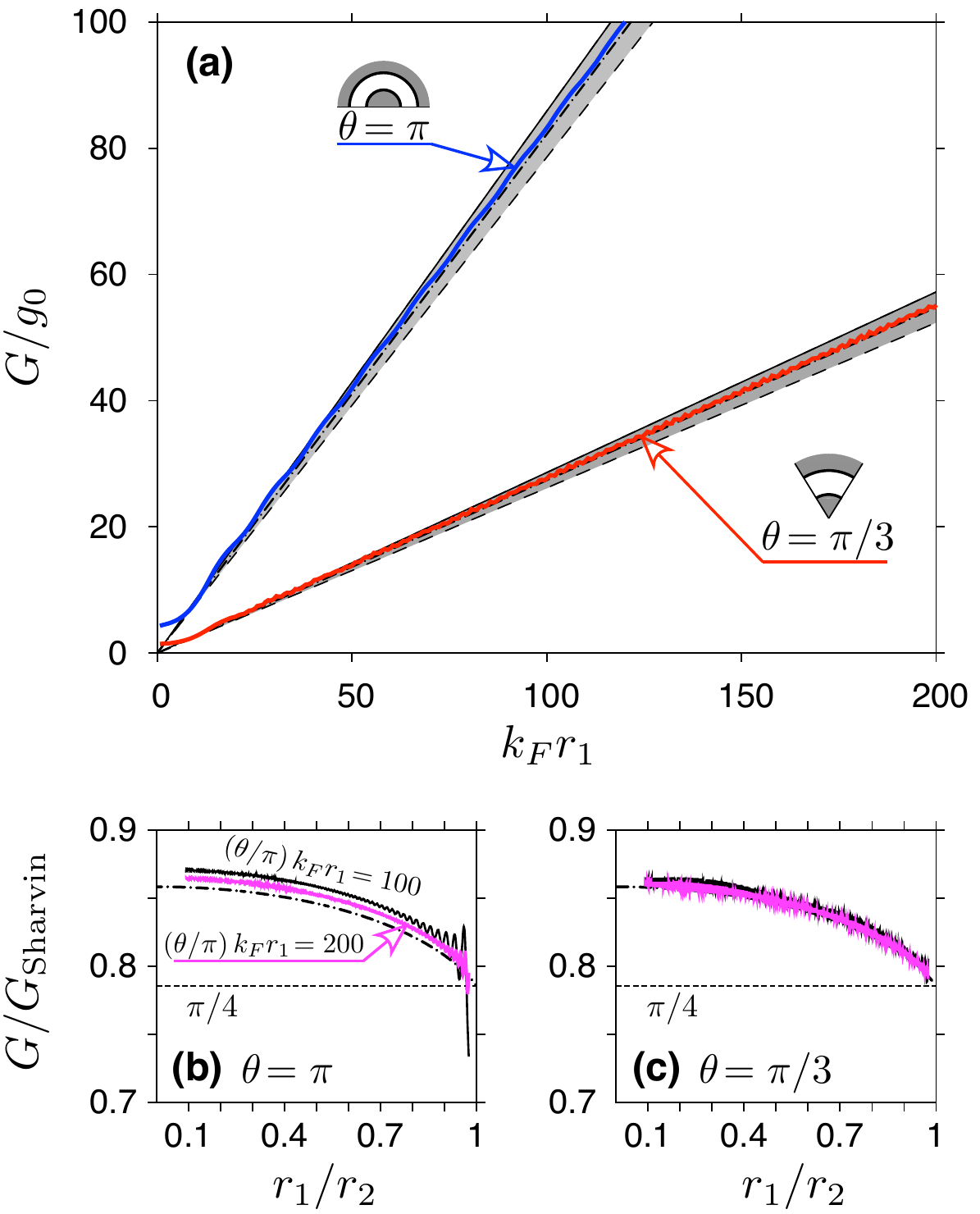}
  \caption{ \label{fig:gshathx}
  (a) Conductance as a~function of the Fermi momentum for the system of
  Fig.\ \ref{fig:setupthx} with the radii ratio $r_1/r_2=0.8$ and the
  opening angles $\theta=\pi$ (blue solid line) and $\theta=\pi/3$
  (red solid line). Shaded areas mark the range between  sub-Sharvin
  conductance $(\pi/4)\,G_{\rm Sharvin}$ [see Eq.\ (\ref{gsharthx})]
  (thin dashed lines) and the value of $(4-\pi)\,G_{\rm Sharvin}$
  (thin solid lines) relevant for the $r_1\ll{}r_2$ limit.
  Dashed-dotted lines correspond $G/G_{\rm Sharvin}\approx{}0.8219$
  obtained from Eq.\ (\ref{ggdiskicoh}).
  (b,c) Solid lines: The conductance reduction as a~function of the radii
  ratio for $\theta=\pi$ and $\theta=\pi/3$ for fixed values of
  $G_{\rm Sharvin}/g_0=100$ and $200$ (same in both panels).
  Dashed-dotted lines mark the approximating formula given by
  Eq.\ (\ref{ggdiskicoh}). The value of $G/G_{\rm Sharvin}=\pi/4$
  is marked with dashed horizontal lines. 
  }
\end{figure}

\begin{figure}[!t]
  \includegraphics[width=0.9\linewidth]{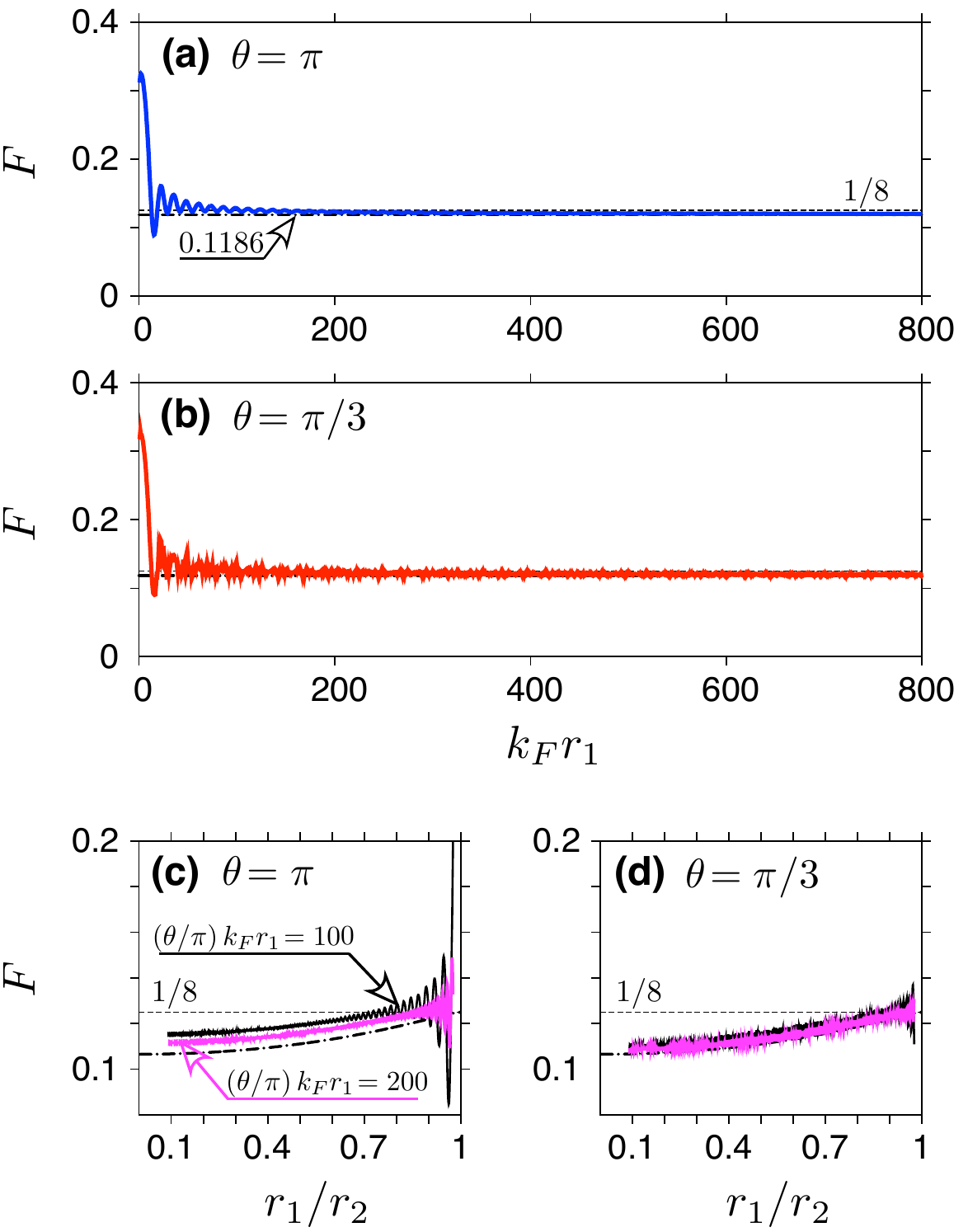}
  \caption{ \label{fig:ffshathx}
  (a,b) Fano factor as a~function of the Fermi momentum for same systems
  as in Fig.\ \ref{fig:gshathx}(a). The opening angle is varied between
  the panels. 
  Horizontal lines show the sub-Sharvin value $F=1/8$
  (dashed) and the value following from Eq.\ (\ref{ffdiskicoh}) for
  $r_1/r_2=0.8$ (dashed-dotted).
  (c,d) Fano factor as a~function of the radii ratio for same values
  of $G_{\rm Sharvin}/g_0$ as used in Figs.\ \ref{fig:gshathx}(b,c).
  Dashed-dotted line depicts the formula given in Eq.\
  (\ref{ffdiskicoh}).
  }
\end{figure}

The conductance and Fano factor obtained by summing over the modes [see
Eq.\ (\ref{gfland}); the limit of $N\rightarrow\infty$ is taken numerically]
are displayed in Figs.\ \ref{fig:gshathx} and \ref{fig:ffshathx}.
The presentation is limited to the two values of the
opening angle, $\theta=\pi$ and $\theta=\pi$, as such examples (together with
the full disk studied in Sec.\ \ref{exadisk}) are sufficient to grasp the
main features introduced with the boundaries. 

In Fig.\ \ref{fig:gshathx}(a) we choose the radii ratio $r_1/r_2=0.8$,
for which the conductance in a~multimode regime (results following from the
mode-matching are depicted with thick solid lines) lays in a~middle of
the range bounded by extreme values following from Eq.\ 
for $r_1/r_2\rightarrow{}1$ and $r_1/r_2\rightarrow{}0$
(see thin dashed and solid lines, respectively), very close to the incoherent
value of $G\approx{}0.8219\,G_{\rm Sharvin}$ (dashed-dotted line) for
$k_Fr_1\gtrsim{}100$.
The corresponding values of the Fano factor in Figs.\ \ref{fig:ffshathx}(a,b)
are closer to $F=1/8$ (the limit of $r_1/r_2\rightarrow{}1$ in Eq.\
(\ref{ffdiskicoh})) even for noticeably higher dopings, however, a~slow
decay towards the value of $F\approx{}0.1186$ following from Eq.\
(\ref{ffdiskicoh}) (dashed-dotted line) is clearly visible.
Also, for both the conductance and the Fano factor displayed as functions
of the radii ratio $r_1/r_2$ for fixed values of $G/G_{\rm Sharvin}=(\theta/\pi)
\,k_F{}r_1$, see (respectively) Figs.\ \ref{fig:gshathx}(b,c) and
Figs.\ \ref{fig:ffshathx}(c,d), we observe a~systematic convergence to the
results following from Eqs.\ (\ref{ggdiskicoh}) and (\ref{ffdiskicoh}),
similarly as for the full disk case in Sec.\ \ref{exadisk}. 

It is worth to notice that a~section of the disk, as depicted in Fig.\
\ref{fig:setupthx}, transforms into a~rectangular sample when
taking the limit of $\theta\rightarrow{}0$ and $r_1/r_2\rightarrow{}1$, such
that the ratio $\theta/(1-r_1/r_2)=\text{const}\equiv{}W/L$.
If additionally the condition for being in a~multimode range, i.e.,
$(\theta/\pi)\,k_F{}r_1\gg{}1$ is satisfied, one can expect, on the basis of
numerical results presented here, that $G/G_{\rm Sharvin}\rightarrow{}\pi/4$
and $F\rightarrow{}1/8$, reproducing the values reported in Ref.\
\cite{Ryc21b}. 

What is more, the scattering in a~disk section bounded with infinite-mass
confinement remain independent for any $j$ channel, with the quantization
given by Eq.\ (\ref{jquanthx}). Therefore, the transmission spectra for
smooth potentials, including the examples shown in Fig.\ \ref{fig:ttkydim2},
will be unaffected and the crossover from the sub-Sharvin to standard Sharvin
transport regime, demonstrated in Sec.\ \ref{smoopots} for the full disk and
in Ref.\ \cite{Ryc21b} for a~rectangle, is predicted to appear also for
a~disk section. 

Although the mathematics required for the mode-matching is a~bit more
cumbersome in the presence of infinite-mass boundaries, we see that key
features of charge transport remain essentially the same as for the full
disk.
An issue not addressed as yet is how the results may be affected by
actual (e.g., {\em irregular}) edges of mesoscopic samples.
Large-scale simulations including possible types of disorder down to an
atomic level are beyond the scope of this work; one should expect,
in analogy with rectangular samples, the consistency between 
theoretical description presented here and experiments to appear for
$\theta/(1-r_1/r_2)\gtrsim{}10$ rather then $\theta/(1-r_1/r_2)\sim{}1$,
i.e., for section of narrow disks with wide opening angles. 
Subsequently, the verification of our predictions for the $r_1\ll{}r_2$ range
may only be possible using the full disk (Corbino) setup.

\section{Conclusions}
\label{conclu}

The effects of sample geometry on selected charge-transfer characteristics
of doped graphene nanosystems have been investigated by comparing the
results for rectangular and disk-shaped (Corbino) setups with different
aspect (or radii) ratios. Finite sections of the disk are also considered. 
Values of the conductance ($G$) and the Fano factor ($F$) obtained from
analytical formulas for transmission probabilities \cite{Two06,Ryc09}
are compared with results
following from the proposed approximating formulas, derived by assuming
incoherent scattering of Dirac fermions between two interfaces separating
weakly- and heavily-doped graphene areas (i.e., electrostatically-doped
sample and the leads).
Numerical analysis of the scattering on a~family of smooth potential
barriers of a~finite height, interpolating between the parabolic and the
rectangular shapes, have also been carried out for the disk, supplementing
our previous study for rectangular samples \cite{Ryc21b}. 

The results show that for rectangular samples the so-called {\em sub-Sharvin}
transport regime, with $G$ being directly proportional to the number of
propagating modes (open channels) and $F\approx{}1/8$, is entered for any
aspect ratio ($W/L$) provided that the doping is sufficiently high, such
that the Fermi wavelength ($\lambda_F$) is much shorter than either the
sample width ($W$) or length ($L$).
Both exact $G$ and $F$ show oscillations (of the Fabry-P\'{e}rot type)
around mean values coinciding with the sub-Sharvin values (derived by
assuming incoherent scattering), with the amplitude decreasing with
increasing $W/L$ or doping.
For disk-shaped samples, as well as for disk sections, 
the oscillations are suppressed, since inner
and outer interfaces are characterized by different curvatures corresponding
to their radii, $r_1<r_2$, and the double-contact analogy no longer applies.  
What is more, $G$ and $F$ become weakly radii-ratio dependent, with high-doping
limits for thin disks ($r_1\approx{}r_2$) approaching the results for
rectangular samples; in the opposite ($r_1\ll{}r_2$) range, the disk $G$
is slightly enhanced (yet still smaller that the Sharvin conductance),
whereas $F$ is slightly suppressed. 
For smooth potentials, transport properties of familiar quantum point
contacts are restored as soon as $\lambda_F$ becomes comparable with
the characteristic length-scale of a~potential jump $\Delta{}r$. 

These findings illustrate how peculiar transmission dependence on incident
angle for weakly-doped/heavily-doped graphene interface (leading, e.g.,
to the Klein tunneling in case of normal incidence) may affect measurable
quantities of mesoscopic graphene samples.
Next to well-known Sharvin transport occurring in various ballistic
structures, and pseudodiffusive charge transport in undoped graphene samples,
one should also expect non-universal (geometry-dependent) reduction of $G$,
by a~factor varying from $\pi/4$ to $4-\pi$, 
comparing to the Sharvin value $G_{\rm Sharvin}$), and amplification of
the shot-noise power (with $F$ between $(9\pi-28)/(12-3\pi)\approx{}0.1065$
and $1/8$), depending on whether one or two interfaces govern the charge
transport.

Since existing experimental works on various systems in graphene report
either the Sharvin conductance, in case a~constriction governing the
transport is distant from sample-lead interfaces \cite{Ter16}, or and
the values of $F\approx{}1/8$ in case of a~rectangular sample with long
parallel interfaces \cite{Dan08,Lai16}, we think it would be beneficial
to confirm experimentally our predictions for an intermediate situation,
i.e., when the transport is ruled by one interface and a~role of the other
is reduced.

{\it Note added.\/}~--- When the work was in principle complete, we become
aware of experimental work on Corbino disk with radii ratio
up to $r_2/r_1\approx{}4.5$ \cite{Kum21}. At low temperatures, conductance
suppression of about $10\%$ (compared to the Sharvin conductance) is observed,
being not far from our prediction for incoherent scattering
[see Eq.\ (\ref{ggdiskicoh})].

\section*{Acknowledgments}
We thank to Shahal Ilani for the correspondence. 
The work was supported by the National Science Centre of Poland (NCN)
via Grant No.\ 2014/14/E/ST3/00256.
Computations were partly performed using the PL-Grid infrastructure.

%%%%%%%%%%%%%%%%%%%%%%%%%%%%%%%%%%%%%%%%%%%%%%%%%%%%%%%%%%%%%%%%%%%%%%%%%%%%%%


\begin{thebibliography}{}
\bibitem{kflenfoo}
The second condition ($k_FL\gg{}1$) is required to guarantee that
the contribution from evanescent modes (with $|k_y|>k_F$) is insignificant
next to the contribution from propagating modes.
\bibitem{kr1maxfoo}
In a~similar plot shown in Ref.\ \cite{Ryc09} (Fig.\ 5) the difference
between the Sharvin and the exact conductance is difficult to notice since
the presentation is limited to $2k_Fr_1\leqslant{}20$.
\bibitem{simfoo}
Numerical integration of Eqs.\ (\ref{phapri},\ref{phbpri})
were performed utilizing a~standard forth-order Runge-Kutta
algorithm. A~spacial step of $5\,$pm was sufficient to keep
the unitarity error, $\mbox{max}\left(|\epsilon_j|\right)<10^{-8}$,
with $\epsilon_j=|r_j|^2+|t_j|^2-1$. 
Summation over the modes [see Eq.\ (\ref{gfland})] was terminated
when $T_j<10^{-6}$.
\bibitem{rmin2foo}
The difference can be expanded as $(r_{\rm min}\!-\!r_c)/r_0=-\xi/2+
{\cal O}(\xi^2)$, with $\xi=r_0{}E/(r_cV_0)\ll{}1$.
\end{thebibliography}
\end{document}